\documentclass[compsoc,conference,a4paper,10pt,times]{IEEEtran}
\IEEEoverridecommandlockouts
\usepackage{cite}
\usepackage{amsmath,amssymb,amsfonts}
\usepackage{graphicx}
\usepackage{textcomp}
\usepackage{xcolor}
\usepackage{times}
\usepackage{bmpsize}
\DeclareMathOperator*{\argmax}{arg\,max}
\DeclareMathOperator*{\argmin}{arg\,min}
\DeclareMathOperator{\stdev}{stdev}
\usepackage{microtype}
\usepackage{graphicx}
\usepackage{booktabs}
\usepackage{hyperref}
\usepackage{algorithm}
\usepackage[noend]{algpseudocode}
\usepackage{tikz}
\usetikzlibrary{calc}
\usetikzlibrary{patterns}
\usetikzlibrary{shapes.multipart, arrows, positioning,shapes.geometric}
\usetikzlibrary{decorations.pathreplacing}
\usetikzlibrary{calc,shapes,decorations,decorations.shapes,positioning,shadows,arrows,decorations.markings,backgrounds,fit}
\usepackage{url}
\usepackage{float}
\usepackage{changepage}
\usepackage{listings}
\usepackage{setspace}
\usepackage{harpoon}
\usepackage{mathrsfs}
\usepackage{multicol}
\usepackage{multirow}
\usepackage{colortbl}
\usepackage{balance}
\usepackage{caption}
\usepackage[labelformat=simple]{subcaption}
\usepackage{lipsum}
\usepackage{pdflscape}
\usepackage{enumitem}
\usepackage{upgreek}
\usepackage[blocks]{authblk}
\usepackage{url}
\usepackage{soul}
\usepackage{siunitx}
\usepackage[normalem]{ulem}

\newcommand{\etal}{\textit{et al.}}

\tikzset{latent/.style={circle,draw,minimum size=20pt}}

\tikzset{sample/.style={rectangle,draw,minimum size=25pt}}

\tikzset{triangle/.style={fill=desertsand, thick, regular polygon, regular polygon sides=3, minimum size=2pt}}
\tikzset{whitetriangle/.style={fill=white, thick, regular polygon, regular polygon sides=3, minimum size=2pt}}
\tikzset{white/.style={fill=white, regular polygon, regular polygon sides=4, minimum size=25pt}}

\tikzset{output/.style={fill = white, draw, thick, regular polygon, regular polygon sides=8, inner sep = -1pt}}

\definecolor{desertsand}{rgb}{0.93, 0.79, 0.69}
\definecolor{ashgrey}{rgb}{0.7, 0.75, 0.71}
\definecolor{bisque}{rgb}{1, 0.95, 0.87}

\definecolor{dkgreen}{rgb}{0,0.6,0}
\definecolor{gray}{rgb}{0.5,0.5,0.5}
\definecolor{mauve}{rgb}{0.58,0,0.82}

\begin{document}

\date{}

\title{Confusing and Detecting ML Adversarial Attacks with Injected Attractors}

\makeatletter
\newcommand\email[2][]%
   {\newaffiltrue\let\AB@blk@and\AB@pand
      \if\relax#1\relax\def\AB@note{\AB@thenote}\else\def\AB@note{\relax}%
        \setcounter{Maxaffil}{0}\fi
      \begingroup
        \let\protect\@unexpandable@protect
        \def\thanks{\protect\thanks}\def\footnote{\protect\footnote}%
        \@temptokena=\expandafter{\AB@authors}%
        {\def\\{\protect\\\protect\Affilfont}\xdef\AB@temp{#2}}%
         \xdef\AB@authors{\the\@temptokena\AB@las\AB@au@str
         \protect\\[\affilsep]\protect\Affilfont\AB@temp}%
         \gdef\AB@las{}\gdef\AB@au@str{}%
        {\def\\{, \ignorespaces}\xdef\AB@temp{#2}}%
        \@temptokena=\expandafter{\AB@affillist}%
        \xdef\AB@affillist{\the\@temptokena \AB@affilsep
          \AB@affilnote{}\protect\Affilfont\AB@temp}%
      \endgroup
       \let\AB@affilsep\AB@affilsepx
}

\author[1]{\bf Jiyi Zhang}
\author[1]{\bf Ee-Chien Chang}
\author[1,2,3,4]{\bf Hwee Kuan Lee}

\affil[1]{School of Computing, National University of Singapore}
\affil[2]{Bioinformatics Institute, A*STAR Singapore}
\affil[3]{Image and Pervasive Access Lab (IPAL), CNRS UMI 2955}
\affil[4]{Singapore Eye Research Institute}
\email{\tt \url{{jzhang93,changec}@comp.nus.edu.sg}, \url{leehk@bii.a-star.edu.sg}}

\maketitle

\begin{abstract}
Many machine learning adversarial attacks find adversarial samples of a victim model ${\mathcal M}$ by following the gradient of some attack objective functions, either explicitly or implicitly. To confuse and detect such attacks, we take the proactive approach that modifies those functions with the goal of misleading the attacks to some local minima, or to some designated regions that can be easily picked up by an analyzer. To achieve this goal, we propose adding a large number of artifacts, which we called {\em attractors}, onto the otherwise smooth function. An attractor is a point in the input space, where samples in its neighborhood have gradient pointing toward it. We observe that decoders of watermarking schemes exhibit properties of attractors and give a generic method that injects attractors from a watermark decoder into the victim model ${\mathcal M}$. This principled approach allows us to leverage on known watermarking schemes for scalability and robustness and provides explainability of the outcomes. Experimental studies show that our method has competitive performance. For instance, for un-targeted attacks on CIFAR-10 dataset, we can reduce the overall attack success rate of DeepFool~\cite{MoosaviDezfooli2016DeepFoolAS} to 1.9\%, whereas known defense LID~\cite{DBLP:journals/corr/abs-1801-02613}, FS~\cite{DBLP:journals/corr/XuEQ17} and MagNet~\cite{DBLP:journals/corr/MengC17} can reduce the rate to 90.8\%, 98.5\% and 78.5\% respectively.
\end{abstract}


\section{Introduction}
\label{sec:intro}
Machine learning models such as deep neural networks are vulnerable to adversarial attacks~\cite{Szegedy2014IntriguingPO} where a small carefully crafted perturbation on the input could lead to a wrong prediction result. As machine learning gains popularity, such vulnerability has brought forward concerns of machine learning adoptions in environments subjected to adversarial influences, such as biometric authentication, fraud detection and autonomous driving.

There are extensive studies on the defenses of neural network models. Many proposed methods, 
in order to detect and recover from adversarial perturbations, make decision based on characteristics of the perturbed adversarial samples. Such characteristics could be implicitly learnt, e.g. through adversarial training~\cite{Goodfellow2015ExplainingAH} and secondary classification~\cite{DBLP:journals/corr/GrosseMP0M17, DBLP:journals/corr/GongWK17,DBLP:journals/corr/abs-1806-00081}, or explicitly determined such as transformation methods~\cite{DBLP:journals/corr/abs-1711-00117, DBLP:journals/corr/LiL16e, DBLP:journals/corr/abs-1812-03411, kou2020enhancing} and principal component analysis (PCA) based methods~\cite{DBLP:journals/corr/HendrycksG16b,DBLP:journals/corr/BhagojiCM17}.

In this paper, we take a different approach. Instead of focusing on the characteristics of adversarial samples, we consider characteristics of the attack process. We view an attack as an optimization process with respect to some attack objective functions, and our goal is to proactively modify such objective functions so as to confuse and mislead the attacker.
\\ \\[-3pt]
\noindent
{\bf \em Attractor.}\ \ 
We propose explicitly injecting {\em attractors} to the classifier. An attractor is a sample  that influences  the attack objective function, so that gradients of the attack objective function in its neighborhood are pointing toward the attractor.   The attractors serve two purposes.  Firstly, they are ``potholes'' and ``bumps'' injected  into  the otherwise smooth slope,  so as to confuse the search process. Secondly, they can trick the search process to some designated regions which make adversarial samples easily detected by an analyzer or even recovered to give correct prediction results.
\begin{figure}[H]
    \centering
    \begin{subfigure}{.5\linewidth}
      \centering
      \includegraphics[width=0.95\linewidth]{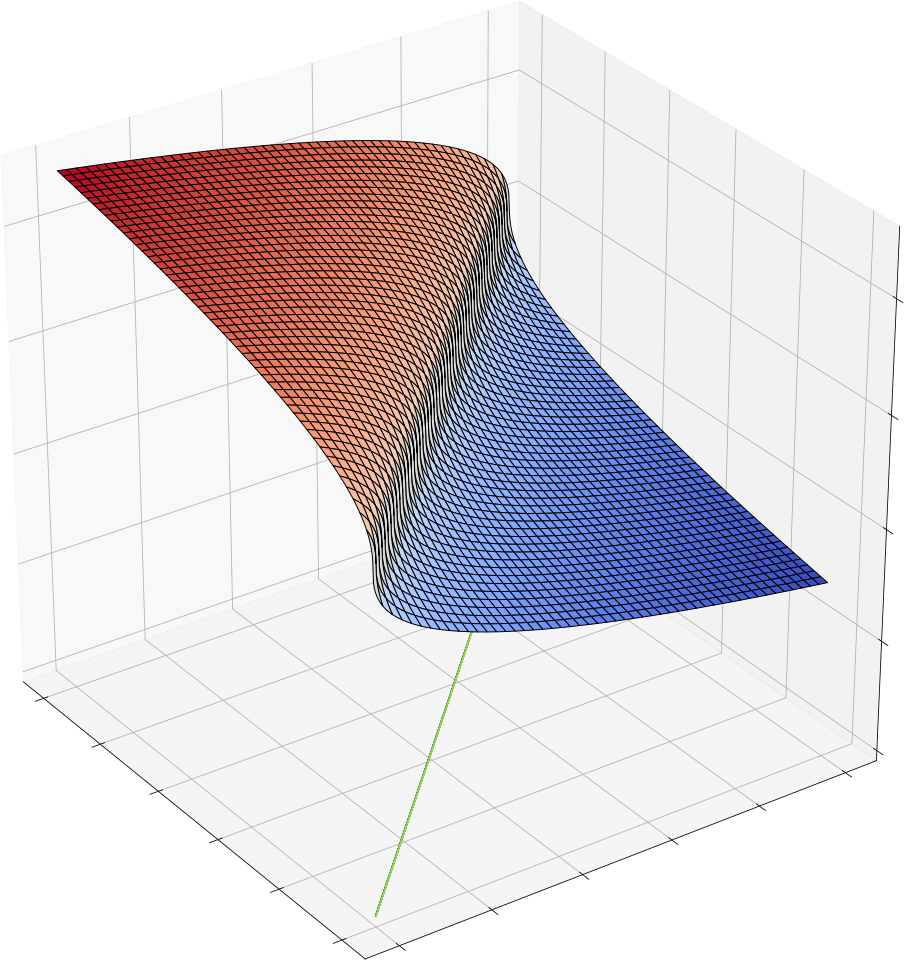}
      \caption{Original victim model}
      \label{fig:1sub1}
    \end{subfigure}%
    \begin{subfigure}{.5\linewidth}
      \centering
      \includegraphics[width=0.95\linewidth]{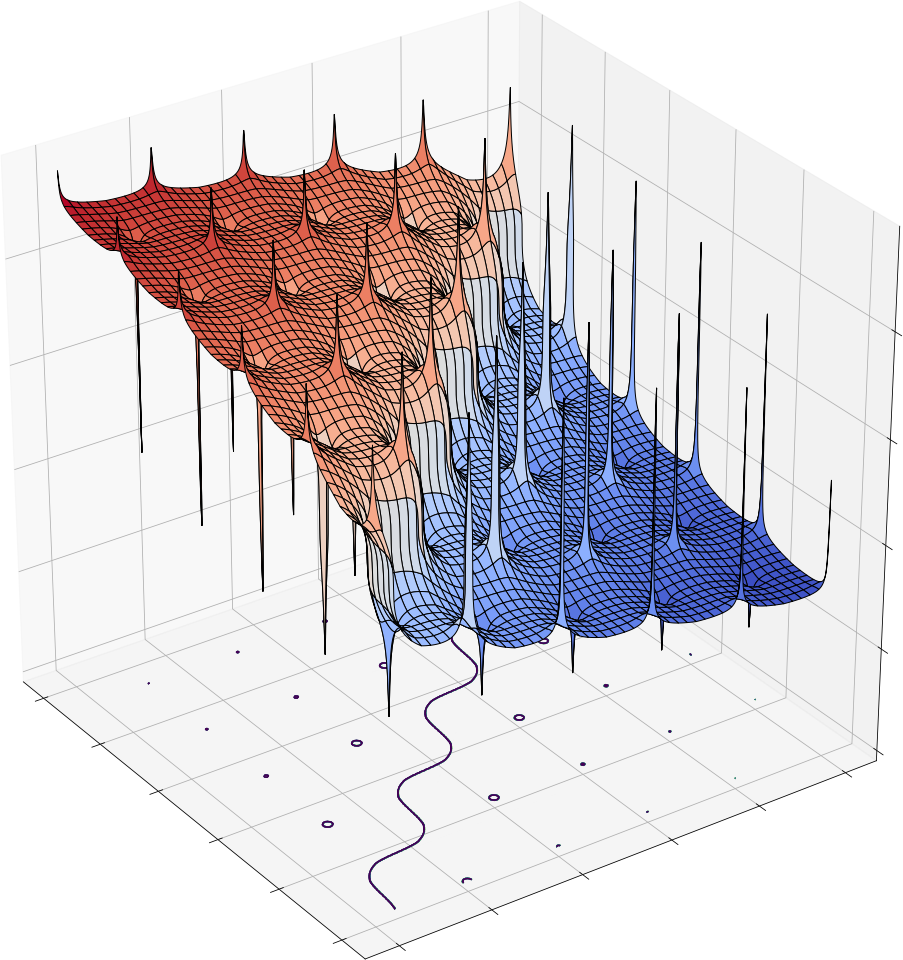}
      \caption{Model with attractors}
      \label{fig:1sub2}
    \end{subfigure}
    \caption{Illustration of attractor. The function of soft label for the class {\tt red}  is shown as a surface.  The classifier's decision boundary (soft label crossing the threshold) is shown on the projected plane.}
    \label{fig:local_minimal}
\end{figure}
Figure~\ref{fig:local_minimal} illustrates attractors on a two-class example. A sample is classified as class {\tt red} if its soft label exceeds certain threshold. The function of the soft label for class {\tt red} is shown as a surface in the figure. Figure~\ref{fig:1sub1} depicts  the original victim classifier  whereas Figure~\ref{fig:1sub2} depicts the situation after attractors are injected. There are two types of attractors: potholes and bumps. Let us consider an attack that  follows the  gradient  so as to minimally perturb a given sample from class  {\tt red}  to class  {\tt blue}. Due to the existence of attractors, this causes the perturbation to move toward a pothole. Similarly, for a sample in class {\tt blue}, a perturbation that follows opposite direction of the gradient will end up in a bump. Samples near potholes and bumps can be designated  and declared as adversarial samples during analysis. With the attractors, attacker would be confused and would either be trapped in some nearby local minima, or move to the designated regions. 

\begin{figure}[H]
    \vspace{-15pt}
    \centering
    \includegraphics[width=0.75\linewidth]{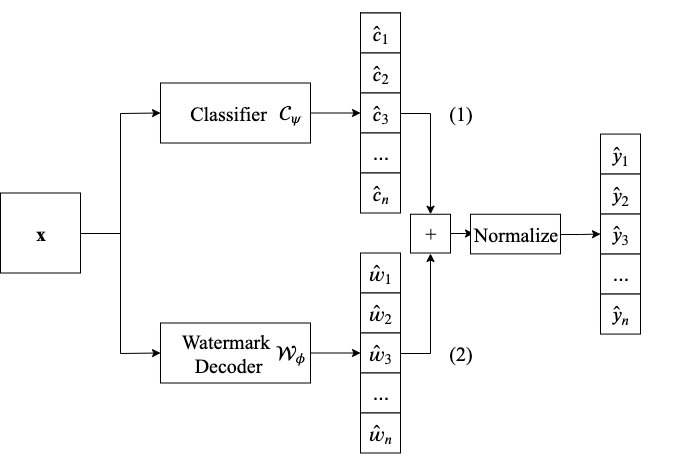}\\
    \text{${\mathcal M}_{(\psi,\phi)}$}
    \caption{The output of the classifier $\mathcal{C}_{\psi}$ and watermark decoder $\mathcal{W}_{\phi}$ are stitched together to form the final output. }
    \label{fig:construction}
\end{figure}

Different attacks optimize different objective functions. As many attacks either directly or indirectly optimize the soft label, a natural candidate of attack objective function is on the soft label. There are also some attacks which do not exploit the soft label, for instance, attacks on black-boxes which only output the final decisions~\cite{DBLP:conf/iclr/BrendelRB18,DBLP:journals/corr/abs-2001-06325, DBLP:conf/iclr/ChengSCC0H20}. To handle such cases, we also consider the ``local density function'' as a candidate which we believe is more generic as it is determined by the classifier's decision boundary.
\\ \\[-3pt]
\noindent
{\bf \em Proposed Construction.}\ \
We propose a construction that injects attractors by adding a flat surface scattered with potholes and bumps onto the classifier's soft labels. Let ${\mathcal C_{\psi}}:\mathcal{X} \rightarrow \mathbb{R}^n$ be the model we want to protect, where $n$ is the number of classes and $\psi$ represents the model parameters. We first choose a robust digital watermarking scheme and its decoder ${\mathcal W_{\phi}}:\mathcal{X} \rightarrow \mathbb{R}^n$, which is coded as a neural network model parameterized by $\phi$, and the $i$-th coefficient of ${\mathcal W_{\phi}}({\textbf x})$ is the correlation value of ${\textbf x}$ with the $i$-th watermark message. One can visualize that the surfaces of ${\mathcal W_{\phi}}({\textbf x})$ are flat but scattered with attractors, since there are watermarked samples everywhere by the fidelity requirement on watermarking schemes. To inject these attractors into the model ${\mathcal C_{\psi}}$,   we simply combine  ${\mathcal C_{\psi}}$ and ${\mathcal W_{\phi}}$ to form a new model $\mathcal{M}_{(\psi, \phi)}$ that outputs the normalized sum of both models' outputs\footnote{It is tempting to adopt some non-linear combination, such as a weighted sum where the weight  varies  with the watermark's strength.   Interestingly, formulation of attractor suggests that such type of weighted sum is a poor choice. In fact, this weighted sum variant is the  counter-example we constructed to show the difference between attractor and trapdoor, a known formulation.}.
In other words, given an input ${\textbf x}$, the final predicted soft labels visible to the adversaries $\hat{y}$ is the normalized 
$ ({\mathcal C_{\psi}} ({\textbf x}) + \mathcal{W}_{\phi}({\textbf x}))$ as illustrated in Figure \ref{fig:construction}.
During analysis, when a sample in question is unusually close to a  particular attractor, we declare it as adversarial.  To see why the method is able to detect and recover, note that by adding the $k$-th soft label,  say $\hat{c}_k$ where $k \in {[1,\dots,n]}$, to the $k$-th  watermark's correlation value $\hat{w}_k$ would ``bind'' the $k$-th class to the $k$-th watermark message.   Now,  if a {\em targeted} attack attempts to increase  the prediction $( \hat{c}_k+\hat{w}_k) $, it would unknowingly increase the correlation to the  corresponding $\hat{w}_k$.       Likewise,  if a {\em un-targeted} attack attempts to decrease the prediction $( \hat{c}_k+\hat{w}_k) $,  the correlation of $\hat{w}_k$ would be decreased. 

Beside conceptually simple, the proposed approach has a few additional advantages.  
The approach is generic and can incorporate different watermarking schemes, and thus can leverage on extensive known works in digital watermarking. For instance, one could employ high capacity watermarking schemes to protect models with large number of classes. Furthermore, when deployed in the black-box setting, no re-training of ${\mathcal C_{\psi}}$ is required. More importantly, the modular approach provides insights into the internal mechanism and provides explainability of the outcomes. 

Nonetheless, a combined model ${\mathcal M}_{(\psi,\phi)}$ obtained by simply stitching ${\mathcal C_{\psi}}$ and ${\mathcal W}_{\phi}$ together is vulnerable to a white-box attack which directly unstitches and obtains ${\mathcal C_{\psi}}$. To prevent such attacks, we can apply neural network obfuscation or distillation techniques \cite{Romero15fitnets:hints,NIPS2015_5784,DBLP:journals/corr/HanMD15,44873,DBLP:journals/corr/abs-1806-10313,DBLP:journals/corr/abs-1901-08121} that transfer the stitched model to the final classifier. In this paper, we propose a method that obfuscates a spread spectrum watermark decoder and a classifier together to demonstrate feasibility of the proposed approach. In addition, the approach of influencing attack process is arguably vulnerable to transfer attacks. Fortunately, known defenses can be combined with attractors. We use an enhanced model to illustrate the combined effectiveness.

We conducted experiments against 18 known attacks and compared our results with some known detection defenses. The results show that our performance is very competitive (results reported in Table \ref{tab:compare_detection}). For example, for un-targeted attack on CIFAR-10 dataset, we can reduce the attack success rate of DeepFool\cite{MoosaviDezfooli2016DeepFoolAS}  to 1.9\%, while known defense   LID\cite{DBLP:journals/corr/abs-1801-02613}, FS\cite{DBLP:journals/corr/XuEQ17}, and MagNet\cite{DBLP:journals/corr/MengC17}
can reduce the rate to  90.8\%, 98.5\% and 78.5\% respectively under the setting where the attack does not utilize the analyzers' outcomes. When the attack utilizes the analyzers' outcomes, our method can reduce to 3.3\%.

Our evaluations are conducted on known attacks. While we are not aware of specific attacks on our method, a potential weakness could be the way various components are stitched/obfuscated together. It would be interesting to further investigate attack and defenses on the stitching method.
\\ \\[-3pt]
\noindent
{\bf \em Related Notions.}\ \
The idea of disrupting the attack process has been explored before. Shumailov \etal~proposed aiding detection by setting up ``Taboo Traps''~\cite{DBLP:journals/corr/abs-1811-07375}, which are  selected neurons  whose activations  are trained to be low on clean images. As the adversaries are not aware of the traps, the generated adversarial samples could  trigger large activation on the taboo traps and thus be identified. Shan \etal~\cite{DBLP:journals/corr/abs-1904-08554} introduced  ``trapdoors'' which are global perturbations that lead to misclassification, and can be viewed as artificial ``shortcuts'' towards misclassification. Since most attacks search for nearby misclassification, the adversarial samples generated  might follow the shortcuts. Similar idea was also explored  in~\cite{li2019invisible} and~\cite{saha2019hidden}. However, the fact that there are misclassification regions nearby does not guarantee that attacks would move toward them. To demonstrate that this concern is legitimate, we give a trapdoor construction that arguably attains the training goals, but yet unable to trick the attackers to follow the trapdoors in Section~\ref{sec:discussion}.  

The notion of attractors is also conceptually relevant to but different from gradient obfuscation methods investigated by Athalye \etal~\cite{athalye2018obfuscated}. A gradient obfuscation method creates a non back-propagatable function so that adversaries are unable to obtain the gradient signal.   In contrast,  gradients of attractors are still smooth and differentiable almost everywhere, and thus could be back-propagatable. In fact,  our intention is to feed gradient information to the attackers so as to pull them nearer to an attractor.
\\ \\
\noindent {\bf \em Contributions:}
\begin{enumerate}[leftmargin=*]
\item We give a formulation of  attractors  and highlight its roles in defending adversarial attacks.  We  point out  crucial difference between this formulation and existing notions, in particular trapdoor.
\item We  propose a generic approach that takes a watermarking decoder and combines  it with the victim model. 
\item We give an obfuscation technique that combines a classifier with spread spectrum watermark decoder.
\item Through extensive experiments against a wide range of attacks and comparison with state-of-art approaches, we demonstrate that the proposed defense attains competitive performance.
\end{enumerate}

\section{Background and Related Works}
\label{sec:background}
\subsection{Attacks}
The goal of adversarial attack is to find a small distortion on the sample that leads to the wrong prediction results, and can be formulated as an optimization problem. To the best of our knowledge, all known efficient attacks utilize some kind of objective functions to guide the searches of adversary samples. Here we categorize them into backpropagation-based attacks\footnotemark~and non-backpropagation\footnotemark[2] based attacks.

\footnotetext{In the literature, they are also knowns as gradient-based and non-gradient based attacks respectively. In our paper, we do not use the term ``gradient'' to address their differences as it may be confusing in this context. Non-backpropagation based attacks can also look for gradients of some functions, even when the soft label is unknown or not differentiable.}
\subsubsection{Backpropagation-based Attacks}
Many attacks make use of the concept of backpropagation. During normal training, a loss between the expected correct output and actual output is computed and backpropagated through the network to compute the gradient of each parameter and then update them accordingly. During attack, adversary fixes parameters but uses an attack objective function to compute a loss and backpropagates it to update the input. Based on the attack objective function, backpropagation-based attacks can be divided into following categories:
\\ \\[-7pt]
\noindent
{\bf \em Derived from Training Loss Function.}\ \
The attack objective function can be derived from the training loss function directly. Such attack objective functions only consider the actual output and adversary's desired output. Fast Gradient Sign Method (FGSM)~\cite{Goodfellow2015ExplainingAH} is a simple and fast approach. The input is perturbed using the direction of backpropagated gradients with a fixed scale $\epsilon$. There are many variations of FGSM which incorporate improvements such as iterative process~\cite{Kurakin2017AdversarialEI,2017arXiv170606083M} and momentum~\cite{DBLP:journals/corr/abs-1710-06081}.
\\ \\[-7pt]
\noindent
{\bf \em Optimization with Less Distortion.}\ \
There are also attack objective functions~\cite{MoosaviDezfooli2016DeepFoolAS,Szegedy2014IntriguingPO, towards,chen2018ead} that do not only compare outputs but also take the amount of perturbation into consideration. The adversary can either add the loss derived from training loss function and amount of perturbation together to optimize them as a whole, or turn the process into a constrained search algorithm which aims to change the prediction result with minimum possible perturbation. Attacks using this type of objective functions are often able to find adversarial samples with smaller perturbation while making some tradeoff in speed of generation.
\\ \\[-7pt]
\noindent
{\bf \em Optimization with Special Goals.}\ \
There are also other attacks using concept of backpropagation but with special attack objective functions. Universal Adversarial Perturbations (UAP)~\cite{DBLP:journals/corr/Moosavi-Dezfooli16} creates image agnostic perturbation that can cause misclassification for most images in a distribution instead of just one image. Jacobian-based Saliency Map Attack (JSMA)~\cite{Papernot2015TheLO} selects only a few pixels in a clean sample and saturates them to cause misclassification.

\subsubsection{Non-backpropagation Based Attacks}
There are attacks which do not rely on backpropagation and use different attack objective functions. Nevertheless, they still can define attack objective functions and use some form of optimization techniques. Such attacks can work in both black-box and white-box settings. Based on their mechanisms, they can be divided into following categories:
\\ \\[-7pt]
\noindent
{\bf \em Score-based Attacks.}\ \
Score-based attacks utilize the soft labels or probability scores provided by the model. Adversary can still compute a loss based on the actual output and desired output. Without backpropagation, non-gradient optimization methods~\cite{DBLP:conf/ccs/ChenZSYH17, DBLP:conf/iclr/RuCBG20, DBLP:journals/corr/abs-1910-02244} can be used to optimize this objective function.
\\ \\[-7pt]
\noindent
{\bf \em Decision-based Attacks.}\ \
When the adversary only has access to the prediction decision without the score, designing an attack becomes more difficult. Boundary attack~\cite{DBLP:conf/iclr/BrendelRB18}, Sign-OPT~\cite{DBLP:conf/iclr/ChengSCC0H20} and attack on attention~\cite{DBLP:journals/corr/abs-2001-06325} belong to this category.
\\ \\[-7pt]
\noindent
{\bf \em Approximation of Gradient.}\ \
There are also scenarios that adversary has full access to the model but cannot get useful gradients due to the existence of some level of defense. Backward Pass Differentiable Approximation (BPDA)~\cite{athalye2018obfuscated} and Simultaneous Perturbation Stochastic Approximation (SPSA)~\cite{pmlr-v80-uesato18a} are designed to approximate gradients for non-differentiable layers to bypass such defenses.

\subsubsection{Transfer-based Attacks}
Adversarial attacks are shown to be transferable in certain way. Papernot~\etal~show that adversarial samples found using a substitute model~\cite{Papernot:2017:PBA:3052973.3053009} could also be adversarial to the original victim model. Adversary can therefore run model extraction~\cite{DBLP:conf/uss/TramerZJRR16} on a black-box victim model, apply known white-box attacks (either backpropagation-based or non-backpropagation based) on extracted model and then transfer adversarial samples to the original black-box model.

\subsubsection{Attacks Used in Our Experiments}
Adversarial machine learning is an active field. We try to cover as many attacks as possible in our experiments. We list all attacks we used for experiments in Appendix~\ref{sec:list_of_attacks}. We refer to Carlini \etal's guideline~\cite{DBLP:journals/corr/abs-1902-06705,DBLP:journals/corr/abs-1905-07112} and consider combinations of different settings.

\subsection{Defenses}
\label{sec:defences}
Many defenses have been proposed to enhance classification robustness based on some difference between the characteristics of clean and adversarial samples.

Characteristics of adversarial samples could be implicitly learnt and used, for example, through adversarial training~\cite{Goodfellow2015ExplainingAH} in which the victim model is trained on adversarial examples and learns to correct them.
They could also be analyzed and identified explicitly. For example, transformation methods~\cite{DBLP:journals/corr/abs-1711-00117, DBLP:journals/corr/LiL16e, DBLP:journals/corr/abs-1812-03411, kou2020enhancing} was proposed based on the characteristic that adversarial samples are not robust against adding noise. Principal component analysis (PCA) based methods~\cite{DBLP:journals/corr/HendrycksG16b,DBLP:journals/corr/BhagojiCM17} is able to recover the correct label by using the characteristic that adversarial perturbation affects mostly the lower-order PCA components.

In our experiments, we compare our proposed method with adversarial training as well as three approaches that explicitly use adversarial characteristics to directly generate a {\tt yes} or {\tt no} answer regarding any given input: Local Intrinsic Dimensionality Based Detector (LID)~\cite{DBLP:journals/corr/abs-1801-02613}, Feature Squeezing Detector (FS)~\cite{DBLP:journals/corr/XuEQ17} and MagNet Detector~\cite{DBLP:journals/corr/MengC17}. We describe these three approaches in Appendix~\ref{sec:list_of_defences}.

\subsubsection{Trapdoor}
Shan~\etal's approach~\cite{DBLP:journals/corr/abs-1904-08554} of using an active method to capture adversarial attacks is related to our idea but with a different motivation. Given a classification model $\mathcal{M}_{\theta}$, a ``trapdoor'' is a global perturbation $\Delta_t$ for a ``trapdoored'' model ${\mathcal{M}_{\theta'}}$ such that $Pr(\argmax(\mathcal{M}_{\theta'}({\textbf x}+ \Delta_t))  = t) \geq 1-\mu$ where $\mu$ is a small constant.

The ``trapdoored'' model ${\mathcal{M}_{\theta'}}$ is obtained through training. The original training dataset is augmented with trapdoor embedded samples. For any ${\textbf x}$ in the training dataset, the label of their trapdoor embedded version ${\textbf x}+\Delta_t$ is set to $t$. The goal of the training is to minimize the classification loss and make the trained model reach an optimal such that it can classify both clean samples and trapdoor embedded samples. As adversarial generation functions naturally gravitate toward these trapdoors, they will produce adversarial samples that the model owner can recognize through a known neuron activation signature. More details will be discussed in Section~\ref{sec:relationship}.
\section{Threat Model}
Our proposed method can be adopted for various attack settings. Let us denote ${\mathcal M}$ as the victim classifier, which is trained to perform certain classification task accurately. Our method injects attractors to ${\mathcal M}$ using a watermark decoder to obtain an ``injected'' classifier $\widehat{\mathcal M}$.
The defense mechanism also includes an additional analyzer, either a detection analyzer ${\mathcal F_d}$ or a recovery analyzer ${\mathcal F_r}$. 

The adversary has access to the $\widehat{\mathcal M}$, ${\mathcal F_d}$ and ${\mathcal F_r}$ with various different capabilities (e.g. adaptive vs non-adaptive,  white-box vs black-box).  
In general, the adversary wants to find a small perturbation $\epsilon$, s.t. the adversarial sample  ${\textbf x} + \epsilon$ is misclassified by $\widehat{\mathcal M}$ to either a specific class (targeted attack) or any other class (non-targeted) attack, while evading detection by the analyzer. The combinations of the adversary’s goal and capability lead to various attack settings. The rest of this section gives details of these settings.

\subsection{Analyzer}
A {\em detection} analyzer is a probabilistic function ${\mathcal F_d}:{\mathcal X} \rightarrow \{{\tt yes},{\tt no}\}$. When given an input ${\textbf x}$,  ${\mathcal F_d}$ decides whether ${\textbf x}$ has been subjected to  attack, and outputs  ${\tt yes}$ iff it deems so.
A {\em recovery} analyzer  ${\mathcal F_r}:{\mathcal X} \rightarrow {\mathcal Y}$ takes a step further.  When given an adversarial sample ${\textbf x}$,  it  outputs a prediction closest to the original prediction prior to the attack.

A defense mechanism could have an explicit analyzer in its workflow, that is, the mechanism first applies classification, and then the analyzer to get the final prediction. The workflow of classify-detect-then-recover can be combined as a single model. On the other hand, some mechanisms do not follow the above mentioned workflow. For example. a model hardened via adversarial training would directly produce the prediction from the input. Nonetheless, there is still an implicit analyzer that is hidden in the inference process.

\subsection{Access to components}
The adversary can have a few levels of access to $\widehat{\mathcal M}$, ${\mathcal F_d}$ and ${\mathcal F_r}$.
\begin{enumerate}
\item[A1.] {\bf {\em No access}}.\ \ This is only applicable to the analyzer. In this case, the adversary is not aware of the existence of the analyzer. While it seems restrictive, this can be adopted in a forensic setting where the input undergoes a check by analyzer before being allowed to proceed to the next stage. 
\item[A2.] {\bf {\em Black-box access}}.\ \ The adversary can feed any input (adaptively) and observe the output.  
\item[A3.] {\bf {\em White-box access}}.\ \ The adversary can see the parameters of the networks. Since the adversary knows the parameters, it can feed the network with any input, observe the internal state and the final output.  
\end{enumerate}
\vspace{5pt}
\noindent
For evaluation purpose, we also consider this setting:
\begin{enumerate}
\item[A2*.] {\bf {\em White-box access without exploiting stitching}}.\ \ Our method stitches a few components together, either with or without obfuscation. 
In this setting, the adversary is given white-box access to a directly stitched and un-obfuscated $\widehat{\mathcal M}$. However, the adversary is not aware of the stitching process. An attack, e.g. FGSM can be applied to the stitched model under white-box, but it does not exploit the stitching. Certainly, it does not make sense to give white-box access to the stitched model in practice, as the adversary can simply extract ${\mathcal M}$ and break the mechanism. Nevertheless, the experimentation result can serve as a good bound on the black-box attack. 
\end{enumerate}
\noindent
Note the adversary does not have access to the original undefended classifier ${\mathcal M}$. Hence, we assume they do not have the training dataset of ${\mathcal M}$, otherwise, they can derive ${\mathcal M}$ from the training dataset. 

\subsection{Targeted/Un-targeted}
When given a clean input ${\textbf x}$, the attacker may have a  specific goal of finding a sample that is being misclassified to 
a particular given class. Such goal is known as {\em targeted attack}.   Alternatively, the attacker could be contended with a weaker  goal that finds a sample that is being misclassified to any class. This is known as {\em un-targeted attack}. To an adversary, un-targeted attack is easier to achieve since the adversarial sample just has to be misclassified.

\subsection{Settings of Attacks}
\label{sec:attack_mode}
In our evaluation, we consider the following combinations:
\\ \\[-7pt]
\noindent
{\bf \em {Non-adaptive Setting}.}\ \ The adversary has black-box access (A2) or white-box access without exploiting stitching (A2*) to the victim classifier $\widehat{\mathcal M}$ and no access (A1) to the analyzer. An attack, e.g. FGSM, would be conducted on the model $\widehat{\mathcal M}$ to obtain the adversarial sample.
Evaluation of many existing detection-based defenses are conducted under non-adaptive setting, where the attacks do not utilize outcomes from the analyzers. For instance, LID, FS, MagNet and Trapdoor are evaluated in this setting. We also evaluate the effectiveness under targeted and un-targeted attack. 
For our evaluation in Section~\ref{sec:non-adaptive}, the injected classifier $\widehat{\mathcal M}$ we use is the stitched model ${\mathcal M}_{(\psi, \phi)}$ constructed by stitching the original classifier with a Quantization Index Modulation (QIM) watermark decoder.
\\ \\[-7pt]
\noindent
{\bf \em {Adaptive Setting}.}\ \ We consider three combinations:
\begin{enumerate}[leftmargin=*]
    \item The adversary has black-box access (A2) or white-box access without exploiting stitching (A2*) to ${\mathcal F_r}$ and $\widehat{\mathcal M}$. In other words, the adversary now can know whether a crafted sample would be accepted by the analyzer. 
    For our evaluation in Section~\ref{sec:attack_on_recovery}, the injected classifier $\widehat{\mathcal M}$ we use is the stitched model ${\mathcal M}_{(\psi, \phi)}$.
    \item The adversary has white-box access (A3) to $\widehat{\mathcal M}$ and black-box access (A2) or white-box access without exploiting stitching (A2*) to ${\mathcal F_r}$. The adversary can investigate the parameters of $\widehat{\mathcal M}$ in order to unstitch and obtain ${\mathcal M}$. 
    We can use obfuscation to prevent such attacks. For our evaluation in Section~\ref{sec:obfs_result}, the injected classifier $\widehat{\mathcal M}$ we use is an obfuscated model $\widetilde{\mathcal{M}}$ with spread spectrum decoder embedded inside.
    \item The adversary has black-box access (A2) or white-box access (A3) to $\widehat{\mathcal M}$ and no access (A1) to the analyzer. Although the attacker does not have the actual training dataset of ${\mathcal M}$, we now suppose it has a similar dataset and some information of the classification task. This implies that the adversary can conduct ``transfer attack''. 
    We can defend such attacks by combining our approach with other existing defenses. For our evaluation in Section~\ref{sec:mitigate_transfer}, we combine our approach with adversary training to construct a model $\widetilde{\mathcal{M}}_{COMB}$ and use it as the injected classifier $\widehat{\mathcal M}$.
\end{enumerate}

\section{Attractors}
\vspace{-5pt}
\subsection{Motivation}
Most  attacks search for adversarial samples along directions derived from  some  local properties. For instance, FGSM takes training loss function's gradient as the search direction. Our goal is to confuse the adversary by adding artifacts to  taint  those local properties, so as to lead the search process to some local minima, or to some designated regions that aid detection.  In a certain sense, the artifacts are potholes and bumps added to an otherwise smooth slope.

This motivates the definition of attractors.   Intuitively,  each attractor ${\textbf x}_0$ is a sample in the input space,  such that the search process would eventually lead all samples in its  neighborhood to ${\textbf x}_0$.  Hence, if the input space is scattered with attractors, then the search process would be confused.  In our formulation,  we call information utilized by the attack as an {\em attack objective function},  and require its gradients pointing  toward the attractors.

Our formulation is  inspired, but different from the attractor in dynamical systems~\cite{ruelle1981}.

\subsection{Definition of Attractors}
\label{sec:definition}
Consider a classification model  $\mathcal{M_\theta}: {\mathcal X} \rightarrow {\mathcal Y}$ parameterized by $\theta$, and let ${\mathcal L}:  
{\mathcal X} \rightarrow {\mathbb R}$ be an attack objective function. 
\\ \\[-7pt]
\noindent {\bf \em Definition.}\ \ We say that a point ${\textbf x}_0 \in {\mathcal X}$ is a {\em $ \mu$-attractor} in  $\mathcal{M}_\theta$ with respect to the attack objective function ${\mathcal L}(\cdot, \cdot,\cdot)$ on ${\textbf y}_t$  if 
 there exists a neighborhood of ${\textbf x}_0$, called the basin of ${\textbf x}_0$ and denoted as $B({\textbf x}_0)$,  such that for all ${\textbf x}\in B({\textbf x}_0)$, 
$$
\cos ( \nabla_{\textbf x} {\mathcal L} ( {\theta, {\textbf x}, {\textbf y}_t}) ,  {\textbf x}_0 -{\textbf x}) \geq 1 - \mu
$$
where  $\cos (\cdot, \cdot)$ is the cosine similarity function
    $ \cos( {\textbf a}, {\textbf b} ) =  \frac{ {\textbf a} \cdot {\textbf b} }{ \left\|  {\textbf a} \right\| \|{\textbf b} \| }$ and ${\textbf y}_t$ is the one-hot vector of the $t$-th class.

\subsection{Choice of Attack Objective Function}
\label{sec:choice_of_objective}
Note that the definition of attractor depends on the attack objective functions. Here are two candidates.
\begin{enumerate}[leftmargin=*]
\item {\bf \em Soft Label.}\ \ Given a model ${\mathcal M}_\theta$, let us choose the attack objective function same as the training loss function, that is ${\mathcal L}(\theta, {\textbf x}, {\textbf y}_t)= J{(\theta, {\textbf x},{\textbf y}_t)}$. Note that   $\nabla_{\textbf x}  J{(\theta, {\textbf x},{\textbf y}_t)}$   is the gradient of the $t$-th soft label of  ${\mathcal M}_{\theta}(\cdot)$ at ${\textbf x}$. This choice of objective function makes sense as many attacks such as FGSM  find the  adversarial sample by moving  along the direction  $\nabla_{\textbf x}  J{(\theta, {\textbf x},{\textbf y}_t)}$. In un-targeted attack with the goal of moving away from class $t$,  the attack moves along the direction  $\nabla_{\textbf x}  J{(\theta, {\textbf x},{\textbf y}_t)}$. Whereas in targeted attack with the goal of finding  a misclassification to class $t$, the attack moves along the direction  $-\nabla_{\textbf x}  J{(\theta, {\textbf x},{\textbf y}_t)}$.
\item {\bf \em Local Density.}\ \ For a model ${\mathcal M}_\theta$ and a sample ${\textbf x}$, let us   define  $(t, \delta)$-{\em local density}, denoted    ${\mathcal H}_{\delta} (\theta, {\textbf x},{\textbf y}_t)$, as the proportion of samples within the sphere of radius $\delta$ centered at ${\textbf x}$ that are classified as the $t$-th class.  That is,
$$   
{\mathcal H}_{\delta} (\theta, {\textbf x}, {\textbf y}_t)  =  \frac{\left| \{ {\widetilde {\textbf x}} \in  S_{\delta}( {\textbf x} ) | \ \argmax({\mathcal M}_\theta   ({\widetilde {\textbf x}})) = t       \}  \right| }{  | S_{\delta}( {\textbf x} ) |}
$$
where $S_{\delta}( {\textbf x} )$ is the sphere of radius $\delta$ centered at ${\textbf x}$.
If ${\textbf x}$ belongs to the   class $t$, we would expect  the local density ${\mathcal H}_{t, \delta} ({\textbf x})$ to be large. Local density, as a choice of objective function for attractors, is more relevant to adversarial attacks that make decision based on the  predicated class instead of the numeric prediction score.
\end{enumerate}
To protect against different attacks,  we look for a model that possesses  attractors with respect to a wide range of attack objective functions. In addition, if a model contains attractors w.r.t. the attack objective function ${\mathcal L}$,  the model should also contain attractors w.r.t. $-{\mathcal L}$.  This is to cater for both targeted attacks and un-targeted attacks which optimize in the opposite directions.

\section {Proposed Method: Attractors from Watermarking}
\subsection {Main Idea}
\label{sec:main_idea}
Our construction makes use of a known model  ${\mathcal W}_\phi$ that exhibits properties of attractors.  To protect a victim classifier ${\mathcal C}_\psi$,  we ``inject'' attractors from ${\mathcal W}_\phi$ into  ${\mathcal C}_\psi$, giving a new model ${\mathcal M}_{{(\psi,\phi)}}$. The new model ${\mathcal M}_{(\psi,\phi)}$ binds each training class of ${\mathcal C}_\psi$ to a class of attractors in  ${\mathcal W}_\phi$, and the binding is achieved by simply giving the normalized sum of  ${\mathcal C}_\psi (\textbf x) +{\mathcal W}_\phi({\textbf x})$ on input ${\textbf x}$.  

We choose  a  watermarking scheme and take its decoder  as ${\mathcal W}_\phi$.  Under a watermarking scheme with capacity of $n$ messages,
each sample in the domain can be decoded to one of the $n$ messages. The goal of the decoder, similar to a classifier, is to determine the message embedded in the sample ${\textbf x}$. In this paper, we treat the decoder as a function  ${\mathcal W}_{\phi}:{\mathcal X} \rightarrow {\mathbb R}^n$, where a coefficient of the output is the correlation value (or confidence level) of the input with a message. Hence the decoded message is  the $i$-th message, where  $i=\argmax {\mathcal W}_\phi({\textbf x})$.  By treating the decoder ${\mathcal W}_{\phi}$ as a function,   the notion of attractors can be applied. 

By summing ${\mathcal W}_{\phi}$  with   ${\mathcal C}_\psi$ as described in the previous paragraph, the $i$-th training class of  ${\mathcal C}_\psi$ is being bound to the $i$-th watermark message.   Hence if an adversary intends to maximize/minimize the $i$-th coefficient  of the sum $( {\mathcal C}_\psi (\cdot) +{\mathcal W}_\phi(\cdot))$, it would unknowingly maximize/minimize the  correlation with the $i$-th watermarking message.

\subsection {Classifier $\mathcal{M}_{(\psi,\phi)}$}
\vspace{-4pt}
Figure \ref{fig:construction} illustrates our construction.  Given the original classifier ${\mathcal C_\psi}$ and the choice of watermark decoder ${\mathcal W}_{\phi}$,  on input ${\textbf x}$, the classifier  $\mathcal{M}_{(\psi,\phi)}$ outputs the normalized sum, i.e. 
\begin{equation}
\label{eq:nsum}
\mathcal{M}_{(\psi,\phi)}({\textbf x}) = \frac{ {\mathcal C_\psi} ({\textbf x}) + {\mathcal W}_{\phi}({\textbf x}) } { | {\mathcal C_\psi} ({\textbf x}) + {\mathcal W}_{\phi}({\textbf x})   |}
\end{equation}
In our experimentation, we choose spread spectrum~\cite{DBLP:journals/tip/CoxKLS97} and Quantization Index Modulation (QIM)~\cite{Chen2001} as the digital watermarking scheme. Section~\ref{sec:qim} gives details of our implementation.

\subsection{Analyzer}
\vspace{-4pt}
\noindent
{\bf \em Detection Analyzer.}\ \
\label{sec:detection}
The {\em detection} analyzer ${\mathcal F_d}:{\mathcal X} \rightarrow \{{\tt yes},{\tt no}\}$ decides whether an input sample is adversarial. 
In our proposed method, on input ${\textbf x}$,  ${\mathcal F_d}$  makes decision based on the values  of $\hat{w}={\mathcal W}_{\phi}({\textbf x})$, and some predefined thresholds $U, L, \sigma_0$.  The detection analyzer declares ${\textbf x}$ as adversarial iff any of the following three conditions holds:
$$C1: \max(\hat{w}){>}U ;\ C2:  \min(\hat{w}){<}L;\ C3:  \stdev(\hat{w}){>}\sigma_0$$
where the vector $\hat{w}$ is treated as a sequence of real values, and  $\mbox{stdev}(\cdot)$ is  the standard deviation over these real values. 
The predefined thresholds are determined by conducting statistical tests on clean and adversarial data.   The above conditions essentially determine whether the coefficients in $\hat{w}$  are anomalies. Unusually high  correlation (condition $C1$) indicates targeted attack. Unusually low correlation (condition $C2$) indicates un-targeted attack. Large variation  (condition $C3$) indicates  optimization and searching has been conducted on the sum. To see that, recap that each watermark is bound to a class as described in Section~\ref{sec:main_idea}.
\\ \\[-7pt]
\noindent
{\bf \em Recovery Analyzer.}\ \  
The {\em recovery} analyzer ${\mathcal F_r}:{\mathcal X} \rightarrow {\mathcal Y}$ is only invoked on input  ${\textbf x}$ that is declared as adversarial by the detection analyzer. On an input ${\textbf x}$ that is declared as un-targeted attack by the detection analyzer, the recovery analyzer returns $\argmin {\mathcal W}_{\phi}(\textbf x)$. For input that is declared as targeted attack, it returns $\argmax {\mathcal C}_{\psi}(\textbf x)$.

To see that why we choose the above logic, note that if the victim ${\textbf x}$ is from $i$-th class,
during an un-targeted attack, the attacker attempts to reduce the $i$-th coefficient of the normalized  sum in equation (\ref{eq:nsum}),   and would unknowingly suppress   the $i$-th coefficient of ${\mathcal W}_{\phi}({\textbf x})$. Therefore, the smallest coefficient is likely the original class. 

During a targeted attack to $j$-th class where $j \not= i$,  the attacker attempts to maximally increase the $j$-th coefficient with some fixed perturbation. It would be harder to recover the original class from ${\mathcal W}_{\phi}({\textbf x})$. Therefore, we can use $\argmax{\mathcal C}_{\psi}(\textbf x)$ to recover the correct class as ${\mathcal C}_{\psi}(\textbf x)$ is likely unaffected. In fact, our analysis in Section~\ref{sec:analysis} shows that due to the confusing effect of attractor, the value change caused by adversarial perturbation is mostly reflected in the output of ${\mathcal W}_{\phi}(\textbf x)$ while ${\mathcal C}_{\psi}({\textbf x})$ is not affected much.

In our proposed analyzers, the outcomes are decided based on explainable logic. It is also possible to construct an analyzer by training a classier that distinguishes adversarial samples using the output of ${\mathcal C}_{\psi}({\textbf x})$ and ${\mathcal W}_{\phi}(\textbf x)$.

\subsection{Watermark Decoder}
\noindent
{\bf \em Spread Spectrum Decoder.}\ \
\label{sec:spread_spectrum}
Let us describe spread spectrum watermarking in the context of image data. Given a $\ell$-pixel image ${\textbf x}=(x_1, x_2, \ldots, x_\ell)$ and a particular $\ell$-bit watermark message  ${\textbf m}=m_1 m_2 \ldots m_\ell$, spread spectrum decoder defines the distance between ${\textbf x}$ and  ${\textbf m}$ as:
 $$  D( {\textbf x}, {\textbf m} )= \sum_{i=1}^\ell    \alpha_i \cdot  (x_i - m_i)^{2}
 $$
where $\alpha_i$ are some predefined weights. The distance (which is low for image containing the message)  is to  be mapped to the correlation value (which is high for image containing the message) by a decreasing function  $S:{\mathbb R} \rightarrow {\mathbb R}$, so that the distances are in the range $[0,1]$. Overall, on input ${\textbf x}$, $\mathcal{W}_{\phi} $ outputs $\langle \hat{w}_1, \hat{w}_2, \ldots, \hat{w}_n\rangle$,  where $$\hat{w}_i = S(D( {\textbf x}, {\textbf m}_i ))$$ for each $i$, where $\langle {\textbf m}_1, \ldots, {\textbf m}_n \rangle$ are $n$ pre-selected $\ell$-bit messages that are embedded/hardcoded into $\mathcal{W}_{\phi}$.

\begin{figure}[H]
    \begin{figure}[H]
        \centering
        \includegraphics[width=0.8\linewidth]{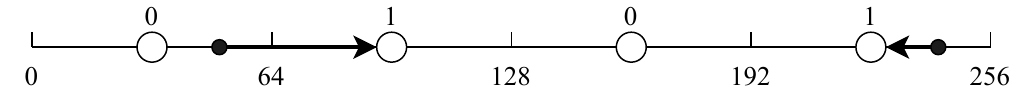}
    \end{figure}
    \vspace{-10pt}
    \caption{An example of quantization on pixel (with step size of 64). The black dots are the pixel values, and the message is {\tt 1}.}
\label{fig:qim_interval}
\end{figure}

We can easily implement the decoder in the form of neural networks.
\\ \\[-7pt]
\noindent
{\bf \em QIM Decoder.}\ \
\label{sec:qim}
We can also adopt a basic variant of  Quantization Index Modulation (QIM) watermarking scheme~\cite{Chen2001}  for $\mathcal{W}_{\phi}$. In the context of image data, the distance  of $\ell$-pixel image ${\textbf x}=(x_1, x_2, \ldots, x_\ell)$ and  a particular $\ell$-bit watermark message  ${\textbf m}=m_1 m_2 \ldots m_\ell$ is determined in the following way:  The value of the $i$-th pixel is quantized with a predefined step size $\delta_i$. The {\em codewords}  for {\tt 0} are at $0.5\delta_i, 2.5\delta_i, \ldots$ and the codewords for {\tt 1} are at  $1.5\delta_i, 3.5\delta_i, \ldots$.    The distance of the $i$-th pixel to $m_i$ is its distance to the nearest codeword for $m_i$ as illustrate in Figure \ref{fig:qim_interval}.  Let us denote the distance as $q(x_i, m_i, \delta_i)$.
\vspace{-10pt}
\noindent
We take the weighted sum of distances over all the pixels as the distance between ${\textbf x}$ and ${\textbf m} $, that is,
$$Q( {\textbf x}, {\textbf m} )= \sum_{i=1}^\ell    \alpha_i \cdot  q(x_i, m_i, \delta_i)$$
where $\alpha_i$ are some predefined weights. Similar to spread spectrum decoder, we map distances to correlation values. On input ${\textbf x}$,  $\mathcal{W}_{\phi} $ outputs $\langle \hat{w}_1, \hat{w}_2, \ldots, \hat{w}_n\rangle$,  where $$ \hat{w}_i = S(  Q( {\textbf x}, {\textbf m}_i ) ) $$ for each $i$, where   $\langle {\textbf m}_1, \ldots, {\textbf m}_n \rangle$ are $n$ pre-selected $\ell$-bit messages that are embedded/hardcoded into $\mathcal{W}_{\phi}$.
\\ \\[-7pt]
\noindent
{\bf \em Remarks.}\ \
There are a few advantages of using a watermarking decoder:   (1) By fidelity requirement of watermarking scheme, the watermarked samples/attractors are scattered over ${\mathcal X} $ and thus for any point in ${\mathcal X} $, there is a nearby watermarked sample/attractor.  (2) There is an efficient decoder to  detect watermarked sample/attractor. (3) There are extensive studies on watermarking in the past two decades, which we can leverage on for construction and analysis of attractors.

\subsection{Obfuscation}
\label{sec:obfuscation}
Given two models ${\mathcal C}_{\psi}$ and ${\mathcal W}_{\phi}$, we want to find a model ${\mathcal M}$ that outputs ${\mathcal C}_{\psi} ({\textbf x}) + {\mathcal W}_{\phi}({\textbf x})$ on input ${\textbf x}$, such that from the layout of ${\mathcal M}$, it is difficult to obtain ${\mathcal C}_{\psi}$ and ${\mathcal W}_{\phi}$. That is, we want to obfuscate the stitched model. In our experiment, we apply the method to spread spectrum decoder.
\\ \\[-7pt]
\noindent
{\bf \em Model Architecture.}\ \
We first select a base neural network structure where the last layer does not have activation function and the last layer contains $n+1$ nodes, where $n$ is the number of classes. Let us denote this structure as $\mathcal{C} = ({\mathcal{S},\mathcal{L}_0})$ where $\mathcal{L}_0$ represents the last layer, including its weights, and $\mathcal{S}$ consists of all the layers before the last. We construct another $\mathcal{W} = ({\mathcal{S},\mathcal{L}_1})$ where the parameters of $\mathcal{S}$ in $\mathcal{C}$ and $\mathcal{W}$ are shared and $\mathcal{L}_1$ has same shape as $\mathcal{L}_0$.
\\ \\[-7pt]
\noindent
{\bf \em Training Procedure.}\ \
We adopt alternate training here: we train ${\mathcal C}$ to classify the training dataset $\mathcal{D}$ and we train ${\mathcal W}$  to classify watermarks. To be specific, when a sample ${\textbf x} \in \mathcal{D}$ is fed into ${\mathcal C}$, it should be classified correctly into one of the classes from $0$ to $n$. When a watermark is fed into ${\mathcal C}$, it will be classified as the $n+1$th class. Vice versa, when a watermark is fed into ${\mathcal W}$, it should be classified with the correct watermark ID from $0$ to $n$. When a sample ${\textbf x} \in \mathcal{D}$ is fed into ${\mathcal W}$, it will be classified as the $n+1$th class.
\\ \\[-7pt]
\noindent
{\bf \em Conversion for Inference.}\ \
When the alternate training is completed, both ${\mathcal C}$ and ${\mathcal W}$  attain high accuracy. We convert them to a final model $\mathcal{M} = ({\mathcal{S},\mathcal{L}_2})$ where $\mathcal{L}_2$ is obtained by summing the corresponding weights in $\mathcal{L}_0$ and $\mathcal{L}_1$ and removing the $n+1$th node. 

After these operations, an attacker will neither be able to simply differentiate this obfuscated model from another ordinary model nor able to undo the obfuscation. At the same time, this model outputs the soft label which is the sum of the classification result and watermark score, thus achieves the effect of attractors.
\\ \\[-7pt]
\noindent
{\bf \em Remarks.}\ \
Note the notion of obfuscation here is different from gradient obfuscation methods investigated by Athalye \etal\cite{athalye2018obfuscated}. A gradient obfuscation method creates a non-backpropagatable function so that adversaries are unable to obtain the gradient signal.   In contrast,  gradients of attractors are still smooth and differentiable almost everywhere even after obfuscation, and thus are backpropagatable. In fact,  our intention is to feed gradient information to the attackers so as to pull them nearer to an attractor.
\section{Evaluation}
\label{sec:eval}
In this section, we first analyze the influences of attractors. We then benchmark our method against some well-known adversarial attacks and compare the results with state-of-art defenses in the common non-adaptive setting. Ling \etal~released DEEPSEC~\cite{Ling2019DEEPSECAU} which is a platform for security analysis of deep learning models. We conduct our experiment using this platform for comparison. We also benchmark our method in adaptive setting.
\subsection{Dataset}
We tested our approach using two datasets: MNIST and CIFAR-10. MNIST contains 60,000 training images and 10,000 testing images. CIFAR-10 contains 50,000 training images and 10,000 testing images. Both of these two datasets have 10 classes. MNIST samples are greyscale images with size of $28 \times 28 \times 1$ and CIFAR-10 samples are colored images with size of $32 \times 32 \times 3$.
\subsection{Model Setup}
We constructed three models for our evaluations:
\begin{itemize}[leftmargin=*]
\item ${\mathcal M}_{(\psi, \phi)}$: The classifier stitched together with a QIM decoder mentioned in Section~\ref{sec:qim}.
\item $\widetilde{\mathcal{M}}$: The obfuscated classifier with spread spectrum decoder embedded inside described in Section~\ref{sec:obfuscation}.
\item $\widetilde{\mathcal{M}}_{COMB}$: A combination of obfuscated classifier $\widetilde{\mathcal{M}}$ and another defense (e.g. adversarial training).
\end{itemize}
We constructed these models for both MNIST and CIFAR-10 datasets. For MNIST dataset, we used a standard CNN as base model. For CIFAR-10, we used ResNet-20 as base model. These models are same as the raw models used in the DEEPSEC. For ${\mathcal M}_{(\psi, \phi)}$ which does not require retraining, its weight is also obtained from DEEPSEC\footnote{\url{https://github.com/kleincup/DEEPSEC}}. For the QIM quantization in ${\mathcal W}_{\phi}$, we represent the pixel as values from [0,255] and use two interval sizes: 3 and 128. The QIM setting is the same in experiments on MNIST and CIFAR-10 datasets.
\subsection{Attractors' Influences}
\label{sec:analysis}
In this section, we investigate (1) how injected attractors affect accuracy (2) whether the proposed method successfully injects attractors. Our analysis is conducted on model ${\mathcal M}_{(\psi, \phi)}$ without considering adversarial attacks.
\\ \\[-7pt]
\noindent
{\bf \em Strength of Victim Model vs Attractors.}\ \
This experiment compares the  ``strength'' of  victim model  against  the watermark decoder on clean input,  so as to verify that they can meet the requirements on classification accuracy and attractors.

\begin{figure}[H]
\centering
\begin{subfigure}{.5\linewidth}
\centering
\vspace{0pt} \includegraphics[width=0.95\linewidth]{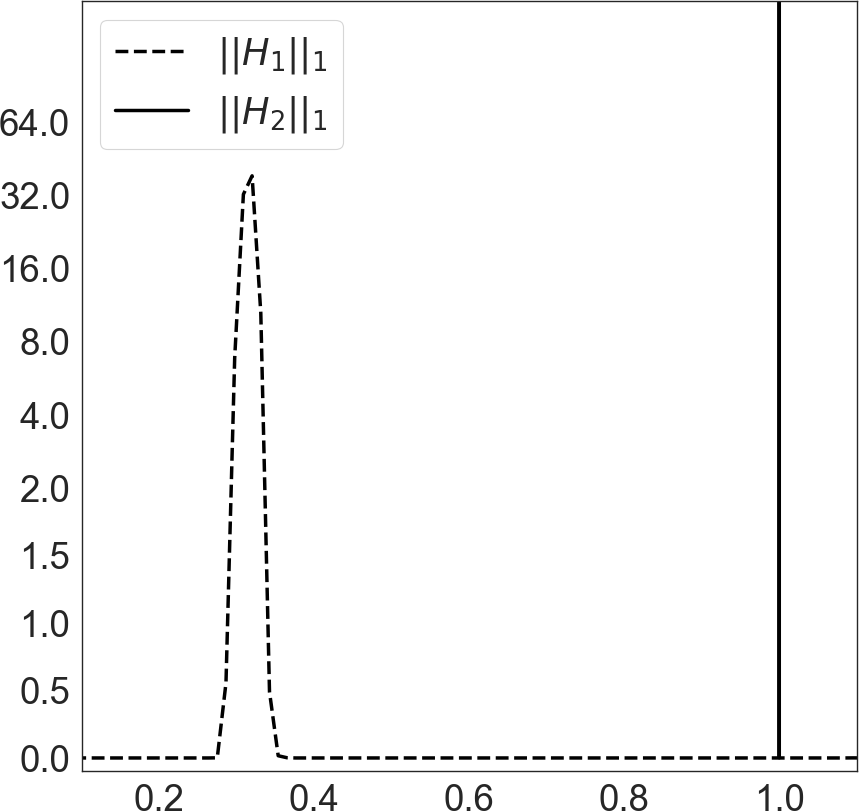}
\caption{}
\label{fig:5sub1}
\end{subfigure}%
\begin{subfigure}{.5\linewidth}
\centering
\vspace{0pt} \includegraphics[width=0.95\linewidth]{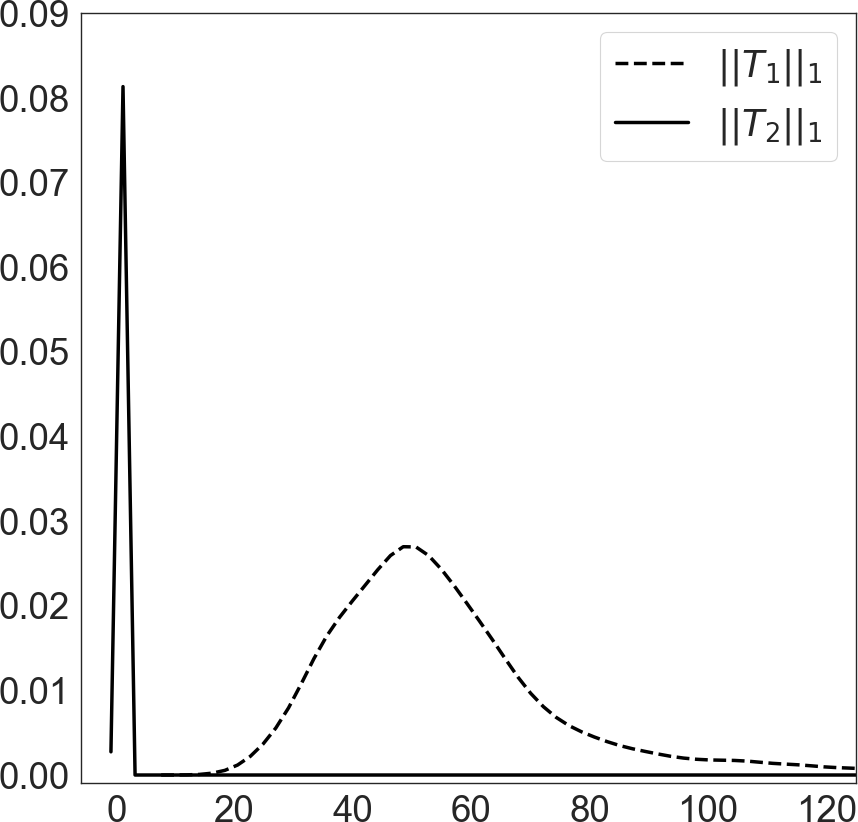}
\caption{}
\label{fig:5sub2}
\end{subfigure}
\caption{Comparing  magnitude (1-norm) of outputs and gradients from ${\mathcal C}_{\psi}$  
and  $\mathcal{W}_{\phi}$. 
(a)  KDE plot of $\|H_1\|_1$ and  $\|H_2\|_1$ where $ H_1 = \mathcal{W}_{\phi}({\textbf x})$ and $ H_2 = {\mathcal C}_{\psi}({\textbf x})$. (b) KDE plot of $\|T_1\|_1$ and  $\|T_2\|_1$ where $ T_1 = \nabla_{{\textbf x}}  J{(\phi, {\textbf x},{\textbf y}_{true})}$ and $T_2 = \nabla_{{\textbf x}}  J{(\psi, {\textbf x},{\textbf y}_{true})}$.}
\label{fig:combined_effect}
\end{figure}

We first feed 10,000 testing images from MNIST dataset into both  the victim model $\mathcal{C}_{\psi}$ and the watermark decoder $\mathcal{W}_{\phi}$. Next, for each testing image,  we  measure the magnitudes (w.r.t. 1-norm) of the output, and magnitudes of the gradients.
In other words, we are comparing the signal at (1) and (2) in Figure~\ref{fig:construction}.
Kernel Density Estimate (KDE) derived from the  measurements are shown in Figure~\ref{fig:combined_effect}. 

Figure~\ref{fig:5sub1} shows that  the magnitude of the output from the original $\mathcal{C}_{\psi}$ is much larger than the magnitude of the output from the decoder $\mathcal{W}_{\phi}$. In other words, in Figure~\ref{fig:construction}, the signal at (1) dominates (2), and thus the attractors have small impact on  the classification accuracy on clean input.

On the other hand, from Figure~\ref{fig:5sub2}, we can see that the gradient's magnitude  from  $\mathcal{W}_{\phi}$ is  much larger than those from  the original model $\mathcal{C}_{\psi}$.  Hence, during attack,  the attacker's optimization strategy would  mostly affected by the watermark decoder instead of the original model.
\noindent
{\bf \em Soft Label.}\ \
\label{sec:softlabel}
This experiment verifies that the direction of gradients from ${\mathcal M}_{(\psi,\phi)}$ indeed exhibit properties of attractors. To verify this property, for each testing image ${\textbf x}$, we find the nearest attractor ${\textbf x}_w$ and its label $t$ and  determine the cosine similarity between $({\textbf x_w}- {\textbf x})$ and the gradient  $-\nabla_{{\textbf x}}  J{\left((\psi, \phi), {\textbf x},{\textbf y}_{t}\right)}$ at ${\textbf x}$.   We also repeat the measurement on  the original model ${\mathcal C}_{\psi}$, that is,  measuring the cosine similarity of  $({\textbf x_w}- {\textbf x})$ and the gradient  $-\nabla_{{\textbf x}}  J{\left(\psi, {\textbf x},{\textbf y}_{t}\right)}$.

\begin{figure}[H]
\centering
\begin{subfigure}{.5\linewidth}
\centering
\vspace{0pt} \includegraphics[width=0.95\linewidth]{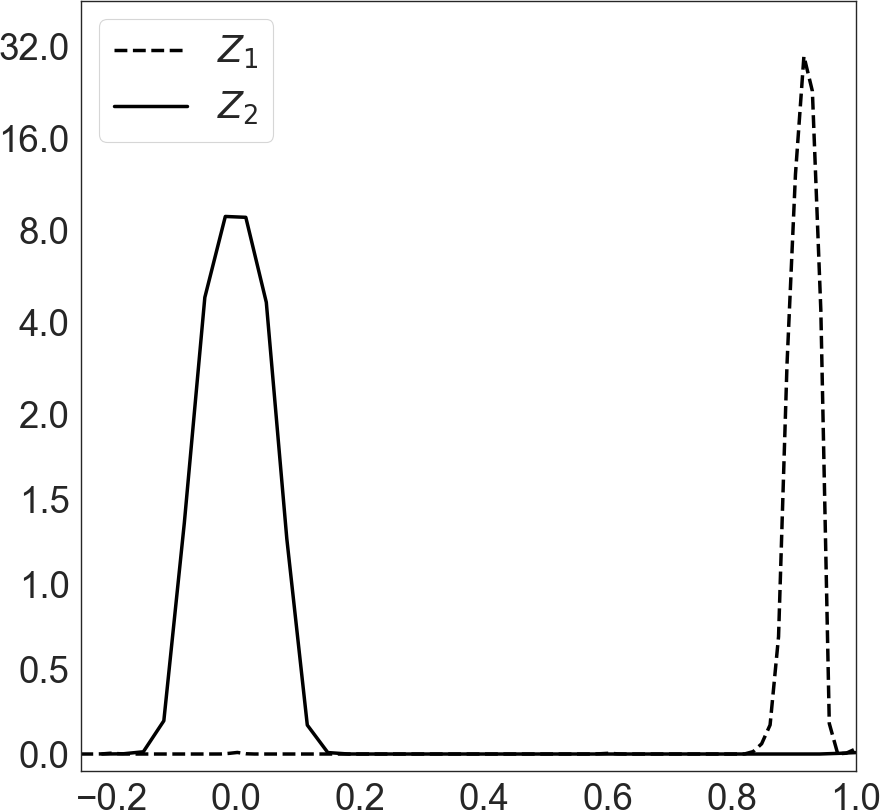}
\caption{}
\label{fig:cosine}
\end{subfigure}%
\begin{subfigure}{.5\linewidth}
\centering
\vspace{0pt} \includegraphics[width=.95\linewidth]{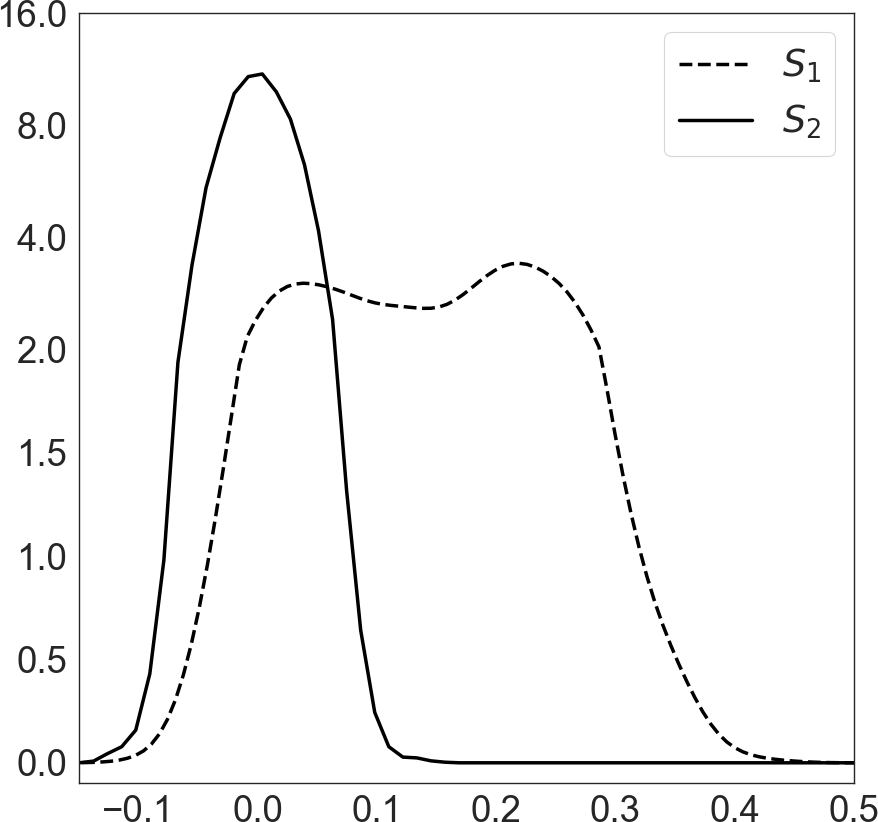}
\caption{}
\label{fig:density}
\end{subfigure}
\caption{(a) KDE plot of  cosine similarity between the direction  to the  nearby attractor $({\textbf x}_w - {\textbf x})$ and the  gradient for a randomly chosen ${\textbf x}$. $Z_1 = \cos(({\textbf x}_w- {\textbf x}),-\nabla_{{\textbf x}}  J{\left(\psi, {\textbf x},{\textbf y}_{t}\right)})$ and $Z_2 = \cos(({\textbf x}_w- {\textbf x}),-\nabla_{{\textbf x}}  J{\left((\psi,\phi), {\textbf x},{\textbf y}_{t}\right)})$. (b) KDE plot of cosine similarity between the direction to the nearby attractor $({\textbf x}_w - {\textbf x})$, and the gradient of local density function for a randomly chosen ${\textbf x}$. $S_1 = \cos(({\textbf x}_w- {\textbf x}),\nabla_{{\textbf x}}  \mathcal{H}_{\delta}{\left(\psi, {\textbf x},{\textbf y}_{t}\right)})$ and $S_2 = \cos(({\textbf x}_w- {\textbf x}),\nabla_{{\textbf x}}  \mathcal{H}_{\delta}{\left((\psi,\phi), {\textbf x},{\textbf y}_{t}\right)})$.}
\end{figure}

The experiment is conducted with 10,000 testing images in MNIST. The KDE of the measurements  are shown in Figure~\ref{fig:cosine}. Note the clear separation between them. Furthermore, note that for ${\mathcal M}_{(\psi,\phi)}$, cosine similarity is more than $0.8$, inferring that a randomly chosen clean sample is likely to have its gradient pointing toward the nearest attractor. Thus we have empirically  verified that the basins of  $\mu$-attractors cover the sample space, where $\mu < 0.2$.
\\ \\
\noindent
{\bf \em Local Density.}\ \
We repeat the experiment described in Section~\ref{sec:softlabel} on local density function and plot the Kernel Density Estimate (KDE) in Figure~\ref{fig:density}. For each testing image ${\textbf x}$, we find the nearest attractor ${\textbf x_w}$ and its label $t$ and determine the cosine similarity between $({\textbf x_w}- {\textbf x})$ and the gradient  $\nabla_{{\textbf x}}  \mathcal{H}_{\delta}{\left((\psi, \phi), {\textbf x},{\textbf y}_{t}\right)}$ at ${\textbf x}$. We also repeat the measurement on the original model $\mathcal{C}_{\psi}$, that is, measuring the cosine similarity of $({\textbf x_w}- {\textbf x})$ and the gradient  $\nabla_{{\textbf x}}  \mathcal{H}_{\delta}{\left(\psi, {\textbf x},{\textbf y}_{t}\right)}$.

The experiment is conducted with 10,000 testing images in MNIST. The KDE of the measurements are shown in Figure~\ref{fig:density}. The result shows that if we choose $\mu =0.9$, a randomly chosen clean sample has more than $50\%$ chance of being in the basin of a $\mu$-attractor.

\subsection{On Clean Data}
\label{sec:clean}
We evaluated the performance of attractors on clean data (10,000 testing images) and reported in Table \ref{tab:onclean}. In this experiment, we use model ${\mathcal M}_{(\psi, \phi)}$ and its analyzers $\mathcal{F}_d$ and $\mathcal{F}_r$. As expected, the classification accuracy of the proposed attractor-embedded model is close to the original victim model ${\mathcal C}_{\psi}$.
\begin{table*}[!t]
    \renewcommand\thetable{2}
    \setlength{\tabcolsep}{1.5pt}
    \footnotesize
    \centering
    \begin{tabular}{|c|c|c|c|c|c|c|c|c|c|c|c|}
    \hline
\multirow{3}{*}{\begin{tabular}[c]{@{}c@{}}(a)\\UA\\/TA\end{tabular}} & \multirow{3}{*}{(b) Attacks} & \multicolumn{5}{c|}{\textbf{MNIST}}                                                                                                                                                                                                                                                                                                                                   & \multicolumn{5}{c|}{\textbf{CIFAR-10}}                                                                                                                                                                                                                                                                                                                                \\ \cline{3-12} 
&                          & \multicolumn{2}{c|}{Attack Success Rate}                                                                                         & \multirow{2}{*}{\begin{tabular}[c]{@{}c@{}}(e)\\ Detection on \\ Misclassfied \\ Input \\ $\mathcal{F}_d$\end{tabular}} & \multirow{2}{*}{\begin{tabular}[c]{@{}c@{}} (f)\\ Recovery on\\ Detected \\ Input \\$\mathcal{F}_r$\end{tabular}} & \multirow{2}{*}{\textbf{\begin{tabular}[c]{@{}c@{}} (g) \\Overall\\ Attack \\ Success\\ Rate\end{tabular}}} & \multicolumn{2}{c|}{Attack Success Rate}                                                                                         & \multirow{2}{*}{\begin{tabular}[c]{@{}c@{}} (j)\\Detection on\\ Misclassfied \\ Input \\$\mathcal{F}_d$\end{tabular}} & \multirow{2}{*}{\begin{tabular}[c]{@{}c@{}}(k)\\Recovery on\\ Detected \\ Input \\$\mathcal{F}_r$\end{tabular}} & \multirow{2}{*}{\textbf{\begin{tabular}[c]{@{}c@{}}($\ell$)\\Overall\\ Attack \\ Success\\ Rate\end{tabular}}} \\ \cline{3-4} \cline{8-9}
&                          & \begin{tabular}[c]{@{}c@{}}(c)\\Original\\ Classifier\\ ${\mathcal C}_{\psi}$\end{tabular} & \begin{tabular}[c]{@{}c@{}}(d)\\With\\ Attractor\\ ${\mathcal M}_{(\psi, \phi)}$\end{tabular} &                                                                        &                                                                       &                                                                                   & \begin{tabular}[c]{@{}c@{}}(h)\\Original\\ Classifier\\ ${\mathcal C}_{\psi}$\end{tabular} & \begin{tabular}[c]{@{}c@{}}(i)\\With\\ Attractor\\ ${\mathcal M}_{(\psi, \phi)}$\end{tabular} &                                                                        &                                                                       &                                                                                   \\ \hline
    \multirow{10}{*}{UA}                     & FGSM     & 80.8\%                                                            & 0.4\%                                                        & 100.0\%                                               & 100.0\%                                              & \textbf{0.0\%}                                                            & 88.7\%                                                            & 53.5\%                                                       & 100.0\%                                               & 99.8\%                                               & \textbf{0.0\%}                                                            \\ \cline{2-12} 
                                                     & RFGSM    &74.2\%                                                            & 0.1\%                                                        & 100.0\%                                               & 100.0\%                                              & \textbf{0.0\%}                                                           & 99.6\%                                                            & 62.7\%                                                       & 100.0\%                                               & 100.0\%                                              & \textbf{0.0\%}                                                            \\ \cline{2-12} 
                                                     & BIM      & 100.0\%                                                                  & 5.6\%                                                        & 100.0\%                                               & 100.0\%                                              & \textbf{0.0\%}                                                            & 100.0\%                                                           & 86.8\%                                                       & 99.9\%                                                & 99.8\%                                               & \textbf{0.1\%}                                                            \\ \cline{2-12} 
                                                     & PGD      & 100.0\%                                                           & 6.9\%                                                        & 100.0\%                                               & 100.0\%                                              & \textbf{0.0\%}                                                            & 100.0\%                                                           & 94.6\%                                                       & 100.0\%                                               & 100.0\%                                              & \textbf{0.0\%}                                                            \\ \cline{2-12} 
                                                     & UMIFGSM  & 100.0\%                                                           & 6.3\%                                                        & 100.0\%                                               & 100.0\%                                              & \textbf{0.0\%}                                                            & 100.0\%                                                           & 100.0\%                                                      & 91.7\%                                                & 98.1\%                                               & \textbf{8.3\%}                                                            \\ \cline{2-12} 
                                                     & UAP      & 39.5\%                                                            & 0.3\%                                                             & 66.7\%                                                      & 100.0\%                                                     &  \textbf{0.1\%}                                                                & 93.1\%                                                            & 87.0\%                                                       & 94.4\%                                                 & 10.9\%                                                & \textbf{4.9\%}                                                           \\ \cline{2-12} 
                                                     & DeepFool & 100.0\%                                                           & 22.3\%                                                       & 100.0\%                                                & 100.0\%                                              & \textbf{0.0\%}                                                           & 100.0\%                                                           & 99.5\%                                                       & 98.1\%                                                & 99.1\%                                               & \textbf{1.9\%}                                                            \\ \cline{2-12} 
                                                     & OM       & 100.0\%                                                           & 96.4\%                                                       & 87.8\%                                                & 100.0\%                                              & \textbf{11.8\%}                                                           & 100.0\%                                                           & 100.0\%                                                             &81.9\%                                                       & 93.8\%                                                     & \textbf{18.1\%}                                                                 \\ \cline{2-12}
                                                     & BPDA       & 100.0\%                                                           & 7.6\%                                                       & 100.0\%                                                & 100.0\%                                              & \textbf{0.0\%}                                                           & 100.0\%                                                           & 92.1\%                                                             &100.0\%                                                       & 100.0\%                                                     & \textbf{0.0\%}                                                                 \\ \cline{2-12}
                                                     & SPSA       & 95.4\%                                                           & 21.1\%                                                       & 100.0\%                                                & 93.7\%                                              & \textbf{0.0\%}                                                           & 90.2\%                                                           & 77.7\%                                                             &100.0\%                                                       & 81.0\%                                                     & \textbf{0.0\%}                                                                 \\ \hline
    \multirow{8}{*}{TA}                     & LLC      & 7.5\%                                                             & 0.0\%                                                        & \multicolumn{2}{c|}{\multirow{3}{*}{No Adversarial Found}}                                                  & \textbf{0.0\%}                                                            & 13.2\%                                                            & 5.4\%                                                        & 98.1\%                                                & 30.2\%                                                     & \textbf{0.1\%}                                                            \\ \cline{2-4} \cline{7-12} 
                                                     & RLLC     & 2.1\%                                                             & 0.0\%                                                        & \multicolumn{2}{c|}{}                                                                                        & \textbf{0.0\%}                                                           & 27.6\%                                                            & 16.0\%                                                       & 100.0\%                                               & 27.7\%                                                     & \textbf{0.0\%}                                                            \\ \cline{2-4} \cline{7-12} 
                                                     & ILLC     & 67.1\%                                                            & 0.0\%                                                        & \multicolumn{2}{c|}{}                                                                                        & \textbf{0.0\%}                                                            & 100.0\%                                                           & 93.7\%                                                       & 97.0\%                                                & 55.7\%                                                     & \textbf{2.8\%}                                                            \\ \cline{2-12} 
                                                     & TMIFGSM  & 84.9\%                                                            & 4.0\%                                                        & 100.0\%                                               &100.0\%                                                      & \textbf{0.0\%}                                                            & 100.0\%                                                           & 100.0\%                                                      & 89.6\%                                                & 22.7\%                                                     & \textbf{10.4\%}                                                           \\ \cline{2-12} 
                                                     & JSMA     & 71.0\%                                                            & 5.3\%                                                        & 89.2\%                                                & 93.1\%                                               & \textbf{0.6\%}                                                            & 100.0\%                                                           & 100.0\%                                                      & 92.9\%                                                 & 10.9\%                                                & \textbf{7.1\%}                                                           \\ \cline{2-12} 
                                                     & BLB      & 100.0\%                                                           & 6.2\%                                                        & 100.0\%                                               & 100.0\%                                              & \textbf{0.0\%}                                                            & 100.0\%                                                           & 100.0\%                                                      & 99.8\%                                                & 61.6\%                                               & \textbf{0.2\%}                                                            \\ \cline{2-12} 
                                                     & CW2      & 100.0\%                                                           & 12.9\%                                                       & 100.0\%                                                & 100.0\%                                              & \textbf{0.0\%}                                                            & 100.0\%                                                           & 100.0\%                                                      & 99.7\%                                                & 97.8\%                                               & \textbf{0.3\%}                                                            \\ \cline{2-12} 
                                                     & EAD      & 100.0\%                                                            & 4.5\%                                                        & 100.0\%                                                & 100.0\%                                              & \textbf{0.0\%}                                                            & 100.0\%                                                           & 77.5\%                                                       & 97.2\%                                                & 95.4\%                                               & \textbf{2.2\%}                                                            \\ \hline
    \end{tabular}
    \caption{Performance of the undefended model and the proposed method against known attacks. Description in Section~\ref{sec:successrate}.}
    \label{tab:attractor_result}
    \vspace{-10pt}
\end{table*}
\begin{table}[H]
\renewcommand\thetable{1}
\footnotesize
\setlength{\tabcolsep}{1.5pt}
\centering
\begin{tabular}{|c|c|c|}
\hline
& \bf{MNIST} & \bf{CIFAR-10} \\ \hline
Victim model ${\mathcal C}_{\psi}$ & 99.3\%         & 90.1\%            \\ \hline
Attractors-embedded model $\mathcal{M}_{(\psi,\phi)}$     & 98.9\%         & 90.0\%            \\ \hline
Attractors-embedded model $\mathcal{M}_{(\psi,\phi)}$ + $\mathcal{F}_d$\footnotemark + $\mathcal{F}_r$     & 98.6\%         & 89.6\%            \\ \hline
\end{tabular}
\caption{Performance of the proposed model $\mathcal{M}_{(\psi,\phi)}$ (for both direct and recovered outputs), and the victim model  ${\mathcal C}_{\psi}$   on clean samples.}
\label{tab:onclean}
\end{table} 
\footnotetext{The false positive rate of $\mathcal{F}_d$ is set to 0.5\% here.}
\subsection{Attack Setup}
We used 18 attacks in total. 10 of them are un-targeted attacks, 8 of them are targeted attacks. Details of these attacks are included in Appendix~\ref{sec:list_of_attacks}.

The details of experiment settings are in Appendix~\ref{sec:settings}.
We use the  settings in DEEPSEC~\cite{Ling2019DEEPSECAU}\footnote{We are aware of the controversy~\cite{DBLP:journals/corr/abs-1905-07112} over DEEPSEC. The results in their paper may give advantages to defender. We use latest version of corrected attacks from GitHub. Among the defenses, we only quote results for detection-only defenses in non-adaptive setting.} to compare our approach with LID, FS and MagNet. The setup is summarized in Table~\ref{table:setup}.  For comparison  with Trapdoor, we use the settings reported by Shan \etal~\cite{DBLP:journals/corr/abs-1904-08554}, which are summarized in Table~\ref{tab:setting_trapdoor}.

\subsection{Evaluation in Non-adaptive Setting}
\label{sec:non-adaptive}
We carried out each of the 18 attacks for both MNIST and CIFAR-10 datasets on model ${\mathcal M}_{(\psi, \phi)}$. Here we are using the attacks in non-adaptive setting to have a fair comparison with existing methods. The attacks here are pit against output of ${\mathcal M}_{(\psi, \phi)}$ directly to cause wrong predictions, without using the output of analyzers.

\subsubsection{Performance of Proposed Method}
\label{sec:successrate}
The results are shown in Table~\ref{tab:attractor_result}. In our experiments, we look at the {\em attack success rate}, {\em detection rate}, {\em recovery rate} and {\em overall attack success rate},  which are described as follows\footnote{We indicated whether the attack is targeted or un-targeted in Column (a) and listed names of the attacks in Column (b). Column (h) to Column (l) repeats the same evaluation as Column (c) to Column (g) on CIFAR-10 dataset.}.

\begin{itemize}[leftmargin=5pt]
\item \textit{\textbf{Attack Success Rate}} measures the performance of a model against attacks. 
We used the testing dataset and both undefended and attractor-embedded models to generate adversarial samples. For each image, if its adversarial sample generated from a model indeed causes that model to make a wrong prediction, it is counted as one success. The attack success rate is measured differently for un-targeted, LLC and targeted attacks. For un-targeted attacks, an attack is counted as successful if the adversarial sample get misclassified into any class other than the correct class. For LLC, an attack is successful if the  adversarial sample is misclassified into the least likely class. For targeted attacks, an attack is successful if the adversarial sample is misclassified into a randomly chosen intended  target class. The {\em attack success rate} on the undefended victim model $\mathcal{C}_{\psi}$, and the attractor-embedded $\mathcal{M}_{(\psi,\phi)} $ are shown in Column (c)  and (d) respectively.
\item \textit{\textbf{Detection Rate}} evaluates detection analyzer ${\mathcal F}_d$'s performance in detecting adversarial behaviors on selected successful adversarial samples. That is, suppose ${\textbf A}$ is the set of  adversarial samples  found by the attack (w.r.t. $\mathcal{M}_{(\psi,\phi)} $),  and ${\textbf B}\subseteq {\textbf A}$ is the set of adversarial samples that being detected by $\mathcal{F}_d$, then the detection rate is $|{\textbf B}|/|{\textbf A}|$. {\em Detection rate} is shown in Column (e).
\item  \textit{\textbf{Recovery Rate}} evaluates recovery analyzer ${\mathcal F}_r$'s performance in recovering correct labels of adversarial samples successfully detected by ${\mathcal F}_d$. Specifically, let ${\textbf B}$ be the set as defined in previous paragraph,  and ${\textbf C}\subseteq {\textbf B}$  the set of adversarial samples that ${\mathcal F}_r$ correctly recover the class,  then the recovery rate is $|{\textbf C}|/|{\textbf B}|$. {\em Recovery rate} is shown in Column (f).
\item \textit{\textbf{Overall Attack Success Rate}} is the percentage of successful and undetected attacks among all  attacks attempts.  Specifically, {\em overall attack success rate}  = {\em attack success rate }$\times$( 1- {\em detection rate}).
Overall attack success rate is a fairer measurement compared with detection rate. To see that, consider the case where a  model $\mathcal{M}_{(\psi,\phi)} $  is  effective in confusing an attack and very few adversarial samples are found by the attack, but the found adversarial samples are the ``difficult'' samples to be detected by  the detector ${\mathcal F}_d$.   In this case, the detection rate is very low, but overall, it is difficult for the attack to find  undetectable  adversarial samples.  In contrast, the overall attack success rate is low in this case, and  fairly reflects  the effectiveness of this attack against the defense methods. {\em Overall attack success rate} is shown in Column (g). 
\end{itemize}
\begin{table*}[!t]
    \scriptsize
    \setlength{\tabcolsep}{1pt}
    \vspace{-18pt}
    \centering
    \begin{tabular}{|c|c|c|c|c|c|c|c|c|c|c|c|c|c|c|c|c|c|c|c|c|c|}
    \hline
    \multicolumn{2}{|c|}{\textbf{MNIST}}                                                         & \multicolumn{4}{c|}{Attractor-embedded Model}                                                                                                              & \multicolumn{4}{c|}{LID}                                                                  & \multicolumn{4}{c|}{FS}                                                                     & \multicolumn{4}{c|}{MagNet}                                                                 & \multicolumn{4}{c|}{Trapdoor}                                                             \\ \cline{1-22} 
    {\begin{tabular}[c]{@{}c@{}}UA\\ /TA\end{tabular}}                & {Attacks}                & \begin{tabular}[c]{@{}c@{}}TPR\\ (a)\end{tabular}                         & \begin{tabular}[c]{@{}c@{}}FPR\\ (b)\end{tabular}                        & \begin{tabular}[c]{@{}c@{}}AUC\\ (c)\end{tabular}                          & \textbf{\begin{tabular}[c]{@{}c@{}}Overall\\Attack \\  Success\\ Rate (d)\end{tabular}} & TPR    & FPR   & AUC    & \textbf{\begin{tabular}[c]{@{}c@{}}Overall\\Attack \\  Success\\ Rate\end{tabular}} & TPR     & FPR   & AUC     & \textbf{\begin{tabular}[c]{@{}c@{}}Overall\\Attack \\  Success\\ Rate\end{tabular}} & TPR     & FPR   & AUC     & \textbf{\begin{tabular}[c]{@{}c@{}}Overall\\Attack \\  Success\\ Rate\end{tabular}} & TPR     & FPR   & AUC   & \textbf{\begin{tabular}[c]{@{}c@{}}Overall\\Attack \\  Success\\ Rate\end{tabular}} \\ \hline
    \multirow{11}{*}{UA}                                              & FGSM                     & 100.0\%                      & 5.0\%                      & 100.0\%                      & \textbf{0.0\%}                                                  & 73.0\% & 3.6\% & 93.7\% & \textbf{8.2\%}                                                  & 96.1\%  & 4.9\% & 99.1\%  & \textbf{1.2\%}                                                  & 100.0\% & 6.6\% & 100.0\% & \textbf{0.0\%}                                                  & 100.0\% & 5.0\% & 100\% & \textbf{0.0\%}                                                  \\ \cline{2-22} 
                                                                      & RFGSM                    & 100.0\%                      & 5.0\%                      & 100.0\%                      & \textbf{0.0\%}                                                  & 70.2\% & 4.1\% & 94.5\% & \textbf{10.2\%}                                                 & 97.7\%  & 3.5\% & 99.5\%  & \textbf{0.8\%}                                                  & 100.0\% & 3.5\% & 100.0\% & \textbf{0.0\%}                                                  & -       & -     & -     & -                                                               \\ \cline{2-22} 
                                                                      & BIM                      & 100.0\%                      & 5.0\%                      & 100.0\%                      & \textbf{0.0\%}                                                  & 10.4\% & 4.2\% & 60.2\% & \textbf{67.7\%}                                                 & 92.7\%  & 3.7\% & 98.7\%  & \textbf{5.5\%}                                                  & 100.0\% & 3.7\% & 100.0\% & \textbf{0.0\%}                                                  & -       & -     & -     & -                                                               \\ \cline{2-22} 
                                                                      & PGD                      & 100.0\%                      & 5.0\%                      & 100.0\%                      & \textbf{0.0\%}                                                  & 10.3\% & 4.1\% & 54.8\% & \textbf{73.9\%}                                                 & 96.1\%  & 3.4\% & 99.5\%  & \textbf{3.2\%}                                                  & 100.0\% & 3.6\% & 100.0\% & \textbf{0.0\%}                                                  & -       & -     & -     & -                                                               \\ \cline{2-22} 
                                                                      & PGD*                     & 100.0\%                      & 5.0\%                      & 100.0\%                      & \textbf{0.0\%}                                                  & -      & -     & -      & -                                                               & -       & -     & -       & -                                                               & -       & -     & -       & -                                                               & 100.0\% & 5.0\% & 100\% & \textbf{0.0\%}                                                  \\ \cline{2-22} 
                                                                      & UMIFGSM                  & 100.0\%                      & 5.0\%                      & 100.0\%                      & \textbf{0.0\%}                                                  & 22.7\% & 4.1\% & 67.6\% & \textbf{54.4\%}                                                 & 90.5\%  & 3.6\% & 98.4\%  & \textbf{6.7\%}                                                  & 100.0\% & 3.7\% & 100.0\% & \textbf{0.0\%}                                                  & -       & -     & -     & -                                                               \\ \cline{2-22} 
                                                                      & UAP                      & 66.7\%                       & 5.0\%                      & 83.3\%                       & \textbf{0.1\%}                                                  & 87.8\% & 4.6\% & 97.5\% & \textbf{3.7\%}                                                  & 99.7\%  & 5.0\% & 99.6\%  & \textbf{0.1\%}                                                  & 100.0\% & 4.0\% & 100.0\% & \textbf{0.0\%}                                                  & -       & -     & -     & -                                                               \\ \cline{2-22} 
                                                                      & DeepFool                 & 100.0\%                       & 5.0\%                      & 100.0\%                       & \textbf{0.0\%}                                                  & 84.1\% & 2.9\% & 98.0\% & \textbf{15.9\%}                                                 & 99.9\%  & 4.0\% & 99.6\%  & \textbf{0.1\%}                                                  & 80.5\%  & 3.6\% & 94.8\%  & \textbf{19.5\%}                                                 & -       & -     & -     & -                                                               \\ \cline{2-22} 
                                                                      & OM                       & 87.8\%                       & 5.0\%                      & 92.9\%                       & \textbf{11.8\%}                                                 & 60.7\% & 3.0\% & 90.0\% & \textbf{39.3\%}                                                 & 94.0\%  & 3.7\% & 99.1\%  & \textbf{6.0\%}                                                  & 91.3\%  & 3.7\% & 97.0\%  & \textbf{8.7\%}                                                  & -       & -     & -     & -                                                               \\ \cline{2-22} 
                                                                      & BPDA                     & 100.0\%                      & 5.0\%                      & \multicolumn{1}{l|}{100.0\%} & \textbf{0.0\%}                                                  & -      & -     & -      & -                                                               & -       & -     & -       & -                                                               & -       & -     & -       & -                                                               & 100.0\% & 5.0\% & 100\% & \textbf{0.0\%}                                                  \\ \cline{2-22} 
                                                                      & SPSA                     & 100.0\%                      & 5.0\%                      & \multicolumn{1}{l|}{100.0\%} & \textbf{0.0\%}                                                  & -      & -     & -      & -                                                               & -       & -     & -       & -                                                               & -       & -     & -       & -                                                               & 100.0\% & 5.0\% & 100\% & \textbf{0.0\%}                                                  \\ \hline
    \multirow{10}{*}{TA}                                              & LLC                      & \multicolumn{3}{c|}{\multirow{3}{*}{No Adversarial Found}}                               & \textbf{0.0\%}                                                  & 87.5\% & 3.6\% & 91.1\% & \textbf{0.7\%}                                                  & 100.0\% & 7.1\% & 99.7\%  & \textbf{0.0\%}                                                  & 100.0\% & 1.8\% & 100.0\% & \textbf{0.0\%}                                                  & -       & -     & -     & -                                                               \\ \cline{2-2} \cline{6-22} 
                                                                      & RLLC                     & \multicolumn{3}{c|}{}                                                                    & \textbf{0.0\%}                                                  & 95.0\% & 5.0\% & 85.3\% & \textbf{0.2\%}                                                  & 100.0\% & 2.5\% & 100.0\% & \textbf{0.0\%}                                                  & 100.0\% & 2.5\% & 100.0\% & \textbf{0.0\%}                                                  & -       & -     & -     & -                                                               \\ \cline{2-2} \cline{6-22} 
                                                                      & ILLC                     & \multicolumn{3}{c|}{}                                                                    & \textbf{0.0\%}                                                  & 64.8\% & 5.9\% & 89.2\% & \textbf{20.9\%}                                                 & 99.7\%  & 3.9\% & 100.0\% & \textbf{0.2\%}                                                  & 100.0\% & 5.2\% & 100.0\% & \textbf{0.0\%}                                                  & -       & -     & -     & -                                                               \\ \cline{2-22} 
                                                                      & TMIFGSM                  & 100.0\%                      & 5.0\%                      & 100.0\%                      & \textbf{0.0\%}                                                  & 52.7\% & 3.5\% & 89.9\% & \textbf{40.9\%}                                                 & 99.3\%  & 3.0\% & 99.9\%  & \textbf{0.6\%}                                                  & 100.0\% & 4.5\% & 100.0\% & \textbf{0.0\%}                                                  & -       & -     & -     & -                                                               \\ \cline{2-22} 
                                                                      & JSMA                     & 89.2\%                       & 5.0\%                      & 92.8\%                       & \textbf{0.6\%}                                                  & 69.1\% & 5.6\% & 92.8\% & \textbf{23.6\%}                                                 & 100.0\% & 3.2\% & 99.6\%  & \textbf{0.0\%}                                                  & 84.0\%  & 5.0\% & 95.3\%  & \textbf{12.2\%}                                                 & -       & -     & -     & -                                                               \\ \cline{2-22} 
                                                                      & BLB                      & 100.0\%                      & 5.0\%                      & 100.0\%                      & \textbf{0.0\%}                                                  & 77.5\% & 5.9\% & 94.7\% & \textbf{22.5\%}                                                 & 99.7\%  & 4.8\% & 99.5\%  & \textbf{0.3\%}                                                  & 98.2\%  & 3.7\% & 99.1\%  & \textbf{1.8\%}                                                  & -       & -     & -     & -                                                               \\ \cline{2-22} 
                                                                      & CW2                      & 100.0\%                       & 5.0\%                      & 100.0\%                       & \textbf{0.0\%}                                                  & 93.9\% & 3.4\% & 99.2\% & \textbf{6.1\%}                                                  & 100.0\% & 3.0\% & 99.6\%  & \textbf{0.0\%}                                                  & 80.5\%  & 3.7\% & 94.5\%  & \textbf{19.4\%}                                                 & -       & -     & -     & -                                                               \\ \cline{2-22} 
                                                                      & CW2*                     & 100.0\%                      & 5.0\%                      & 100.0\%                      & \textbf{0.0\%}                                                  & -      & -     & -      & -                                                               & -       & -     & -       & -                                                               & -       & -     & -       & -                                                               & 97.2\%  & 5.0\% & 99\%  & -                                                               \\ \cline{2-22} 
                                                                      & EAD                      & 100.0\%                       & 5.0\%                      & 100.0\%                       & \textbf{0.0\%}                                                  & 92.0\% & 3.5\% & 98.5\% & \textbf{8.0\%}                                                  & 100.0\% & 3.5\% & 99.4\%  & \textbf{0.0\%}                                                  & 75.8\%  & 4.4\% & 92.3\%  & \textbf{24.2\%}                                                 & -       & -     & -     & -                                                               \\ \cline{2-22} 
                                                                      & EAD*                     & 100.0\%                      & 5.0\%                      & 100.0\%                      & \textbf{0.0\%}                                                  & -      & -     & -      & -                                                               & -       & -     & -       & -                                                               & -       & -     & -       & -                                                               & 98.0\%  & 5.0\% & 99\%  & -                                                               \\ \hline
    \end{tabular}

    \vspace{5pt}
        \begin{tabular}{|c|c|c|c|c|c|c|c|c|c|c|c|c|c|c|c|c|c|c|c|c|c|}
    \hline
    \multicolumn{2}{|c|}{\textbf{CIFAR-10}}                                                         & \multicolumn{4}{c|}{Attractor-embedded Model}                                                                                                              & \multicolumn{4}{c|}{LID}                                                                  & \multicolumn{4}{c|}{FS}                                                                     & \multicolumn{4}{c|}{MagNet}                                                                 & \multicolumn{4}{c|}{Trapdoor}                                                             \\ \cline{1-22} 
    {\begin{tabular}[c]{@{}c@{}}UA\\ /TA\end{tabular}}                & {Attacks}                & \begin{tabular}[c]{@{}c@{}}TPR\\ (a)\end{tabular}                         & \begin{tabular}[c]{@{}c@{}}FPR\\ (b)\end{tabular}                        & \begin{tabular}[c]{@{}c@{}}AUC\\ (c)\end{tabular}                          & \textbf{\begin{tabular}[c]{@{}c@{}}Overall\\Attack \\  Success\\ Rate (d)\end{tabular}} & TPR    & FPR   & AUC    & \textbf{\begin{tabular}[c]{@{}c@{}}Overall\\Attack \\  Success\\ Rate\end{tabular}} & TPR     & FPR   & AUC     & \textbf{\begin{tabular}[c]{@{}c@{}}Overall\\Attack \\  Success\\ Rate\end{tabular}} & TPR     & FPR   & AUC     & \textbf{\begin{tabular}[c]{@{}c@{}}Overall\\Attack \\  Success\\ Rate\end{tabular}} & TPR     & FPR   & AUC   & \textbf{\begin{tabular}[c]{@{}c@{}}Overall\\Attack \\  Success\\ Rate\end{tabular}} \\ \hline
    \multirow{11}{*}{UA}                                              & FGSM                     & 100.0\% & 5.0\%                      & 99.9\%                       & \textbf{0.0\%}                                                  & 100.0\% & 5.1\% & 100.0\% & \textbf{0.0\%}                                                  & 9.5\%  & 2.9\% & 82.6\% & \textbf{81.2\%}                                                 & 99.1\%  & 4.7\% & 93.5\% & \textbf{0.8\%}                                                  & 100.0\% & 5.0\% & 100.0\% & \textbf{0.0\%}                                                  \\ \cline{2-22} 
                                                                      & RFGSM                    & 100.0\% & 5.0\%                      & 100.0\%                      & \textbf{0.0\%}                                                  & 100.0\% & 2.9\% & 100.0\% & \textbf{0.0\%}                                                  & 6.0\%  & 4.8\% & 70.7\% & \textbf{78.7\%}                                                 & 33.3\%  & 3.2\% & 83.2\% & \textbf{55.8\%}                                                 & -       & -     & -     & -                                                               \\ \cline{2-22} 
                                                                      & BIM                      & 99.9\%  & 5.0\%                      & 100.0\%                      & \textbf{0.1\%}                                                  & 94.6\%  & 2.9\% & 99.1\%  & \textbf{5.4\%}                                                  & 1.6\%  & 4.5\% & 25.5\% & \textbf{98.4\%}                                                 & 1.8\%   & 4.2\% & 53.0\% & \textbf{98.2\%}                                                 & -       & -     & -     & -                                                               \\ \cline{2-22} 
                                                                      & PGD                      & 100.0\% & 5.0\%                      & 100.0\%                      & \textbf{0.0\%}                                                  & 99.9\%  & 3.5\% & 100.0\% & \textbf{0.1\%}                                                  & 0.4\%  & 3.8\% & 16.5\% & \textbf{99.6\%}                                                 & 3.2\%   & 4.3\% & 59.2\% & \textbf{96.8\%}                                                 & -       & -     & -     & -                                                               \\ \cline{2-22} 
                                                                      & PGD*                     & 100.0\% & 5.0\%                      & 100.0\%                      & \textbf{0.0\%}                                                  & -       & -     & -       & -                                                               & -      & -     & -      & -                                                               & -       & -     & -      & -                                                               & 100.0\% & 5.0\% & 100.0\% & \textbf{0.0\%}                                                  \\ \cline{2-22} 
                                                                      & UMIFGSM                  & 91.7\%  & 5.0\%                      & 97.7\%                       & \textbf{8.3\%}                                                  & 100.0\% & 3.0\% & 100.0\% & \textbf{0.0\%}                                                  & 1.8\%  & 4.1\% & 23.8\% & \textbf{98.2\%}                                                 & 6.3\%   & 4.1\% & 57.1\% & \textbf{93.7\%}                                                 & -       & -     & -     & -                                                               \\ \cline{2-22} 
                                                                      & UAP                      & 94.4\%   & 5.0\%                      & 97.2\%                       & \textbf{4.9\%}                                                 & 100.0\% & 5.3\% & 100.0\% & \textbf{0.0\%}                                                  & 2.9\%  & 3.8\% & 76.3\% & \textbf{82.8\%}                                                 & 99.5\%  & 5.9\% & 94.9\% & \textbf{0.4\%}                                                  & -       & -     & -     & -                                                               \\ \cline{2-22} 
                                                                      & DeepFool                 & 98.1\%  & 5.0\%                      & 98.6\%                       & \textbf{1.9\%}                                                  & 9.2\%   & 5.7\% & 64.0\%  & \textbf{90.8\%}                                                 & 1.5\%  & 3.9\% & 86.3\% & \textbf{98.5\%}                                                 & 21.5\%  & 2.8\% & 81.0\% & \textbf{78.5\%}                                                 & -       & -     & -     & -                                                               \\ \cline{2-22} 
                                                                      & OM                       & 81.9\%  & 5.0\%                      & 88.3\%                       & \textbf{18.1\%}                                                 & 8.8\%   & 4.9\% & 65.1\%  & \textbf{91.2\%}                                                 & 25.0\% & 3.8\% & 89.0\% & \textbf{75.0\%}                                                 & 46.4\%  & 3.9\% & 78.7\% & \textbf{53.6\%}                                                 & -       & -     & -     & -                                                               \\ \cline{2-22} 
                                                                      & BPDA                     & 100.0\% & 5.0\%                      & \multicolumn{1}{l|}{100.0\%} & \textbf{0.0\%}                                                  & -       & -     & -       & -                                                               & -      & -     & -      & -                                                               & -       & -     & -      & -                                                               & 100.0\% & 5.0\% & 100.0\% & \textbf{0.0\%}                                                  \\ \cline{2-22} 
                                                                      & SPSA                     & 100.0\% & 5.0\%                      & \multicolumn{1}{l|}{100.0\%} & \textbf{0.0\%}                                                  & -       & -     & -       & -                                                               & -      & -     & -      & -                                                               & -       & -     & -      & -                                                               & 100.0\% & 5.0\% & 100.0\% & \textbf{0.0\%}                                                  \\ \hline
    \multirow{10}{*}{TA}                                              & LLC                      & 98.1\%  & 5.0\%                      & 99.9\%                       & \textbf{0.1\%}                                                  & 100.0\% & 1.5\% & 100.0\% & \textbf{0.0\%}                                                  & 3.7\%  & 9.0\% & 73.5\% & \textbf{12.9\%}                                                 & 100.0\% & 6.7\% & 91.8\% & \textbf{0.0\%}                                                  & -       & -     & -     & -                                                               \\ \cline{2-22} 
                                                                      & RLLC                     & 100.0\% & 5.0\%                      & 100.0\%                      & \textbf{0.0\%}                                                  & 99.0\%  & 5.7\% & 99.2\%  & \textbf{0.3\%}                                                  & 11.7\% & 5.1\% & 71.0\% & \textbf{27.8\%}                                                 & 31.4\%  & 3.8\% & 81.2\% & \textbf{21.6\%}                                                 & -       & -     & -     & -                                                               \\ \cline{2-22} 
                                                                      & ILLC                     & 97.0\%  & 5.0\%                      & 97.8\%                       & \textbf{2.8\%}                                                  & 79.2\%  & 5.3\% & 96.1\%  & \textbf{20.8\%}                                                 & 51.7\% & 3.3\% & 83.9\% & \textbf{48.3\%}                                                 & 2.6\%   & 4.7\% & 61.2\% & \textbf{97.4\%}                                                 & -       & -     & -     & -                                                               \\ \cline{2-22} 
                                                                      & TMIFGSM                  & 89.6\%  & 5.0\%                      & 95.2\%                       & \textbf{10.4\%}                                                 & 100.0\% & 5.8\% & 100.0\% & \textbf{0.0\%}                                                  & 10.0\% & 3.8\% & 45.0\% & \textbf{90.0\%}                                                 & 10.4\%  & 3.8\% & 57.9\% & \textbf{89.6\%}                                                 & -       & -     & -     & -                                                               \\ \cline{2-22} 
                                                                      & JSMA                     & 92.9\%   & 5.0\%                      & 97.2\%                       & \textbf{7.1\%}                                                 & 71.5\%  & 3.4\% & 94.4\%  & \textbf{28.4\%}                                                 & 20.6\% & 3.7\% & 91.7\% & \textbf{79.2\%}                                                 & 53.2\%  & 5.3\% & 92.3\% & \textbf{46.7\%}                                                 & -       & -     & -     & -                                                               \\ \cline{2-22} 
                                                                      & BLB                      & 99.8\%  & 5.0\%                      & 99.9\%                       & \textbf{0.2\%}                                                  & 13.0\%  & 3.1\% & 72.3\%  & \textbf{87.0\%}                                                 & 1.7\%  & 4.1\% & 89.3\% & \textbf{98.3\%}                                                 & 52.5\%  & 4.3\% & 81.6\% & \textbf{47.5\%}                                                 & -       & -     & -     & -                                                               \\ \cline{2-22} 
                                                                      & CW2                      & 99.7\%  & 5.0\%                      & 99.8\%                       & \textbf{0.3\%}                                                  & 19.9\%  & 3.8\% & 77.6\%  & \textbf{80.1\%}                                                 & 0.9\%  & 3.7\% & 88.1\% & \textbf{99.1\%}                                                 & 38.4\%  & 4.4\% & 81.8\% & \textbf{61.6\%}                                                 & -       & -     & -     & -                                                               \\ \cline{2-22} 
                                                                      & CW2*                     & 100.0\% & 5.0\%                      & 100.0\%                      & \textbf{0.0\%}                                                  & -       & -     & -       & -                                                               & -      & -     & -      & -                                                               & -       & -     & -      & -                                                               & 96.2\%  & 5.0\% & 97.0\%  & -                                                               \\ \cline{2-22} 
                                                                      & EAD                      & 97.2\%  & 5.0\%                      & 98.3\%                       & \textbf{2.2\%}                                                  & 17.2\%  & 4.0\% & 73.8\%  & \textbf{82.8\%}                                                 & 1.9\%  & 3.5\% & 89.8\% & \textbf{98.1\%}                                                 & 54.2\%  & 5.0\% & 82.1\% & \textbf{45.8\%}                                                 & -       & -     & -     & -                                                               \\ \cline{2-22} 
                                                                      & EAD*                     & 100.0\% & 5.0\%                      & 100.0\%                      & \textbf{0.0\%}                                                  & -       & -     & -       & -                                                               & -      & -     & -      & -                                                               & -       & -     & -      & -                                                               & 95.0\%  & 5.0\% & 97.0\%  & -                                                               \\ \hline
    \end{tabular}
    \caption{We compare detection rate of attractors with LID, FS, MagNet and Trapdoor. Description in Section~\ref{sec:table3}.}
    \label{tab:compare_detection}
    \vspace{-10pt}
    \end{table*}
\noindent
From the results, we have made following observations:
\\ \\[-7pt]
\noindent
{\bf \em Observation 1: Effectiveness of Confusing} \\
Our experiments show that the successful rate on attractor-embedded model $\mathcal{M}_{(\psi,\phi)} $ is significantly lower  when compared with the undefended  model  $\mathcal{C}_{\psi}$. This observation is reflected in  Table~\ref{tab:attractor_result} column (c), (d), (h) and (i). When adversary follows the direction of gradient provided by the { attractors}, they move toward the nearby attractor instead of the   decision boundary, 
and thus may be stuck in a local minimum or incur a larger perturbation, which confuses the attacking process and leads to a lower attack success rate.
\\ \\[-7pt]
\noindent
{\bf \em Observation 2: Effectiveness on Non-backpropagation Attacks}  \\
We achieve high detection accuracy for most backpropagation-based attacks. Note that non-backpropagation based attacks BPDA and SPSA are also not effective against our defense. Although BPDA is successful on defenses that break or hide the gradient, our defense uses gradient to deceive the adversary instead of creating non-backpropagatable functions, and therefore able to trick  BPDA. Similarly, although SPSA uses non-gradient based optimization, taking random small steps indirectly uses  information on the soft label's  gradient, and would still converge to
the nearby attractor.
\\ \\[-7pt]
\noindent
{\bf \em Observation 3: Attractors of Multi-Scale Gradients} \\
The attacks MI-FGSM, JSMA and UAP indirectly carry out some forms of gradient averaging  in deciding the perturbation:
MI-FGSM uses the gradient of previous iterations to avoid falling into the local minimum, JSMA saturates only a few pixels based on the saliency map, and  UAP searches for a universal perturbation through averaging,  that can be applied to most of the samples. In a certain sense, such attacks are making decision based on gradient at a  lower scale  in the  multi-scale gradient representation.
\\ \\[-10pt]
\indent
Hence, to address such attacks, we should have a mixture of attractors catering for attack-loss functions
at different scales. Our implementation achieves this by controlling  the interval size $\delta_i$ and the weightage $\alpha_i$ for each pixel, where a larger $\delta_i$ corresponds to lower scale, and a larger weights corresponds to larger emphasis on the corresponding  scale. In our experiment, we use two interval sizes, 3 and 128, and give more weightage  on larger interval size. 
Empirically, this choice achieves good performance. 
It would be interesting to find an analytical approach to determine the scale.

\subsubsection{Comparison with Known Defenses}
\label{sec:table3}
We compared the performance of our approach with three existing detection analyzers: LID, FS. MagNet as well as Trapdoor in the same setting. The results are shown in Table~\ref{tab:compare_detection}. Here we use {\em true positive rate (TPR)}, {\em false positive rate (FPR)}, {\em area under the curve (AUC)} and {\em overall attack success rate} to evaluate and compare.
\begin{itemize}[leftmargin=5pt]
\item \textit{\textbf{True Positive Rate (TPR)}} is the detection rate on successful adversarial input that get misclassified. TPR is shown in Column (a).
\item \textit{\textbf{False Positive Rate (FPR)}} is the percentage of clean samples which get wrongly detected as adversarial for all samples in the test dataset. FPR is shown in Column (b).
\item \textit{\textbf{Area Under the Curve (AUC)}} computes the area under ROC curve. It is a threshold-independent benchmark for the detection performance. AUC is shown in Column (c).
\item \textit{\textbf{Overall Attack Success Rate}} is defined in the previous Subsection~\ref{sec:successrate} and shown in Column (d). 
\end{itemize}
We adopt and follow the definitions of TPR, FPR and AUC in  DEEPSEC~\cite{Ling2019DEEPSECAU}.
The performance of known mechanisms in Table~\ref{tab:compare_detection} are obtained from results reported in  DEEPSEC~\cite{Ling2019DEEPSECAU} and Trapdoor~\cite{DBLP:journals/corr/abs-1904-08554}. 

Note that in our experiments, a same threshold value was selected to detect all attacks. With this value, FPR could be fixed. On the other hand, a defense can be very effective on certain attacks. For these attacks, there exist better thresholds which can clearly separate clean and adversarial samples, resulting in 100\% AUC.

For PGD, CW2 and EAD, experiments are carried out in two different settings where one setting is same as DEEPSEC, and the other from Shan \etal~\cite{DBLP:journals/corr/abs-1904-08554}\footnote{The data on TPR, FPR and AUC are directly quoted. The data on overall attack success rate are derived from the reported performance. Since data on the initial attack success rate for Trapdoor are not available, we are unable to derive some of the  overall attack success rate for Trapdoor.}. 

\begin{table}[H]
    \vspace{-6pt}
    \centering
    \scriptsize
    \setlength{\tabcolsep}{0.5pt}
    \begin{tabular}{|c|c|c|c|c|c|c|c|c|c|}
        \hline
        &Dataset & \multicolumn{4}{c|}{\textbf{MNIST}}                                                                                                                                                 & \multicolumn{4}{c|}{\textbf{CIFAR-10}}                                                                                                                                              \\ \cline{2-10} 
& FPR                 & \multicolumn{2}{c|}{5\% FPR}                                                             & \multicolumn{2}{c|}{0.5\% FPR}                                                           & \multicolumn{2}{c|}{5\% FPR}                                                             & \multicolumn{2}{c|}{0.5\% FPR}                                                           \\ \hline 
\begin{tabular}[c]{@{}c@{}}UA\\ /TA\end{tabular}    & Attacks                  & TPR & \textbf{\begin{tabular}[c]{@{}c@{}}Overall\\ Attack\\ Success\\ Rate\end{tabular}} & TPR & \textbf{\begin{tabular}[c]{@{}c@{}}Overall\\ Attack\\ Success\\ Rate\end{tabular}} & TPR & \textbf{\begin{tabular}[c]{@{}c@{}}Overall\\ Attack\\ Success\\ Rate\end{tabular}} & TPR & \textbf{\begin{tabular}[c]{@{}c@{}}Overall\\ Attack\\ Success\\ Rate\end{tabular}} \\ \hline

        \multirow{10}{*}{UA}     & FGSM         & 100.0\%                                                                   & \textbf{0.0\%}                                                           & 100.0\%                                                                   & \textbf{0.0\%}                                                           & 100.0\%                                                     & \textbf{0.0\%}                                                           & 100.0\%                                                      & \textbf{0.0\%}                                                           \\ \cline{2-10} 
                                 & RFGSM        & 100.0\%                                                                   & \textbf{0.0\%}                                                           & 100.0\%                                                                   & \textbf{0.0\%}                                                           & 100.0\%                                                     & \textbf{0.0\%}                                                           & 100.0\%                                                      & \textbf{0.0\%}                                                           \\ \cline{2-10} 
                                 & BIM          & 100.0\%                                                                   & \textbf{0.0\%}                                                           & 100.0\%                                                                   & \textbf{0.0\%}                                                           & 99.9\%                                                      & \textbf{0.1\%}                                                           & 99.9\%                                                       & \textbf{0.1\%}                                                           \\ \cline{2-10} 
                                 & PGD          & 100.0\%                                                                   & \textbf{0.0\%}                                                           & 100.0\%                                                                   & \textbf{0.0\%}                                                           & 100.0\%                                                     & \textbf{0.0\%}                                                           & 99.9\%                                                       & \textbf{0.1\%}                                                           \\ \cline{2-10} 
                                 & UMIFGSM      & 100.0\%                                                                   & \textbf{0.0\%}                                                           & 100.0\%                                                                   & \textbf{0.0\%}                                                           & 91.7\%                                                      & \textbf{8.3\%}                                                           & 86.3\%                                                       & \textbf{13.7\%}                                                          \\ \cline{2-10} 
                                 & UAP          & 66.7\%                                                                    & \textbf{0.1\%}                                                           & 66.7\%                                                                    & \textbf{0.1\%}                                                           & 94.4\%                                                      & \textbf{4.9\%}                                                           & 93.6\%                                                       & \textbf{5.6\%}                                                           \\ \cline{2-10} 
                                 & DeepFool     & 100.0\%                                                                   & \textbf{0.0\%}                                                           & 100.0\%                                                                   & \textbf{0.0\%}                                                           & 98.1\%                                                      & \textbf{1.9\%}                                                           & 98.0\%                                                       & \textbf{2.0\%}                                                           \\ \cline{2-10} 
                                 & OM           & 87.8\%                                                                    & \textbf{11.8\%}                                                          & 85.7\%                                                                    & \textbf{13.8\%}                                                          & 81.9\%                                                      & \textbf{18.1\%}                                                          & 81.1\%                                                       & \textbf{18.9\%}                                                          \\ \cline{2-10} 
                                 & BPDA         & 100.0\%                                                                   & \textbf{0.0\%}                                                           & 100.0\%                                                                   & \textbf{0.0\%}                                                           & 100.0\%                                                     & \textbf{0.0\%}                                                           & 99.9\%                                                       & \textbf{0.1\%}                                                           \\ \cline{2-10} 
                                 & SPSA         & 100.0\%                                                                   & \textbf{0.0\%}                                                           & 100.0\%                                                                   & \textbf{0.0\%}                                                           & 100.0\%                                                     & \textbf{0.0\%}                                                           & 100.0\%                                                      & \textbf{0.0\%}                                                           \\ \hline
        \multirow{8}{*}{TA}      & LLC          & \multirow{3}{*}{\begin{tabular}[c]{@{}c@{}}No\\ Adv\\ Found\end{tabular}} & \textbf{0.0\%}                                                           & \multirow{3}{*}{\begin{tabular}[c]{@{}c@{}}No\\ Adv\\ Found\end{tabular}} & \textbf{0.0\%}                                                           & 98.1\%                                                      & \textbf{0.1\%}                                                           & 98.1\%                                                       & \textbf{0.1\%}                                                           \\ \cline{2-2} \cline{4-4} \cline{6-10} 
                                 & RLLC         &                                                                           & \textbf{0.0\%}                                                           &                                                                           & \textbf{0.0\%}                                                           & 100.0\%                                                     & \textbf{0.0\%}                                                           & 98.1\%                                                       & \textbf{0.3\%}                                                           \\ \cline{2-2} \cline{4-4} \cline{6-10} 
                                 & ILLC         &                                                                           & \textbf{0.0\%}                                                           &                                                                           & \textbf{0.0\%}                                                           & 97.0\%                                                      & \textbf{2.8\%}                                                           & 90.7\%                                                       & \textbf{8.7\%}                                                           \\ \cline{2-10} 
                                 & TMIFGSM      & 100.0\%                                                                   & \textbf{0.0\%}                                                           & 100.0\%                                                                   & \textbf{0.0\%}                                                           & 89.6\%                                                      & \textbf{10.4\%}                                                          & 81.1\%                                                       & \textbf{18.9\%}                                                          \\ \cline{2-10} 
                                 & JSMA         & 89.2\%                                                                    & \textbf{0.6\%}                                                           & 88.0\%                                                                    & \textbf{0.6\%}                                                           & 92.9\%                                                      & \textbf{7.1\%}                                                           & 90.1\%                                                       & \textbf{9.9\%}                                                           \\ \cline{2-10} 
                                 & BLB          & 100.0\%                                                                   & \textbf{0.0\%}                                                           & 100.0\%                                                                   & \textbf{0.0\%}                                                           & 99.8\%                                                      & \textbf{0.2\%}                                                           & 99.6\%                                                       & \textbf{0.4\%}                                                           \\ \cline{2-10} 
                                 & CW2          & 100.0\%                                                                   & \textbf{0.0\%}                                                           & 100.0\%                                                                   & \textbf{0.0\%}                                                           & 99.7\%                                                      & \textbf{0.3\%}                                                           & 99.7\%                                                       & \textbf{0.3\%}                                                           \\ \cline{2-10} 
                                 & EAD          & 100.0\%                                                                    & \textbf{0.0\%}                                                           & 100.0\%                                                                    & \textbf{0.0\%}                                                           & 97.2\%                                                      & \textbf{2.2\%}                                                           & 97.2\%                                                       & \textbf{2.2\%}                                                           \\ \hline
        \end{tabular}
    \caption{Detection performance at 0.5\% FPR.}
    \vspace{-8pt}
\label{fig:lowfpr}
\end{table}
\subsubsection{Performance at Low False Positive Rate}
\label{sec:lowfpr}
Our experiments follow settings in DEEPSEC where the false positive rates (FPR) of LID, FS, MagNet as well as Trapdoor are set to be around  $5\%$.  However, there are application scenarios where a $5\%$ FPR would generate too many false alarms and not acceptable. We conduct another experiment on our proposed method where the FPR is adjusted to $0.5\%$, and the results are shown in Table~\ref{fig:lowfpr}. Note that the proposed method still achieves good performance.

\subsection{Evaluation in Adaptive Setting}
\vspace{-10pt}
In this subsection, we evaluate the performance of our approach under three adaptive attack settings.
\vspace{-2pt} 
\subsubsection{Attack on Recovery Analyzer}
\label{sec:attack_on_recovery}
In this experiment, we use the same victim ${\mathcal M}_{(\psi, \phi)}$ and assume that the adversary has access to the recovered class label produced by the analyzers. In this setting, the attacks are pit against output of the recovery analyzer to cause wrong predictions and evade the analysis at same time. 

We also tested adversarial training in the same setting for comparison. The results are shown in Table \ref{tab:merged}. For unbounded attacks such as BLB, CW2 and EAD, we set maximum number of iterations according to Table~\ref{table:setup} and measure the $L_2$ distortion of the successful adversary samples in Table~\ref{tab:l2}.  Note that the successful samples generated on our model has larger $L_2$ distortion than on the original classifier. 
\begin{table}[H]
    \scriptsize
    \setlength{\tabcolsep}{0.0pt}
    \centering
    \begin{tabular}{|c|c|c|c|c|c|c|c|} 
    \hline
                                                       & Dataset  & \multicolumn{3}{c|}{\textbf{MNIST} }                                                                                                                                                                                                              & \multicolumn{3}{c|}{\textbf{CIFAR-10} }                                                                                                                                                                                                            \\ 
    \hline
    \begin{tabular}[c]{@{}c@{}} UA\\ /TA \end{tabular} & Attacks  & \begin{tabular}[c]{@{}c@{}}Original\\ Classifier\\ ${\mathcal C}_{\psi}$ \end{tabular} & \begin{tabular}[c]{@{}c@{}}Adversarial\\Training\end{tabular} & \begin{tabular}[c]{@{}c@{}}With\\ Attractor\\ ${\mathcal M}_{(\psi, \phi)}$ \end{tabular} & \begin{tabular}[c]{@{}c@{}}Original\\ Classifier\\ ${\mathcal C}_{\psi}$ \end{tabular} & \begin{tabular}[c]{@{}c@{}}Adversarial\\Training\end{tabular} & \begin{tabular}[c]{@{}c@{}}With\\ Attractor\\ ${\mathcal M}_{(\psi, \phi)}$ \end{tabular}  \\ 
    \hline
    \multirow{10}{*}{UA}                               & FGSM     & 80.8\%                                                                                 & 35.0\%                                                         & 0.5\%                                                                                    & 88.7\%                                                                                 & 22.2\%                                                        & 4.9\%                                                                                     \\ 
    \cline{2-8}
                                                       & RFGSM    & 74.2\%                                                                                 & 23.5\%                                                         & 0.0\%                                                                                    & 99.6\%                                                                                 & 81.9\%                                                        & 5.3\%                                                                                     \\ 
    \cline{2-8}
                                                       & BIM      & 100\%                                                                                  & 90.0\%                                                         & 0.6\%                                                                                    & 100.0\%                                                                                & 100.0\%                                                        & 5.0\%                                                                                     \\ 
    \cline{2-8}
                                                       & PGD      & 100\%                                                                                  & 90.3\%                                                         & 0.0\%                                                                                    & 100.0\%                                                                                & 100.0\%                                                        & 6.3\%                                                                                     \\ 
    \cline{2-8}
                                                       & UMIFGSM  & 100\%                                                                                  & 92.6\%                                                         & 0.7\%                                                                                    & 100.0\%                                                                                & 100.0\%                                                        & 8.6\%                                                                                     \\ 
    \cline{2-8}
                                                       & UAP      & 39.5\%                                                                                 & 1.5\%                                                             & 14.1\%                                                                                   & 93.1\%                                                                                 & 73.8\%                                                        & 28.1\%                                                                                    \\ 
    \cline{2-8}
                                                       & DeepFool & 100.0\%                                                                                & 90.9\%                                                         & 2.5\%                                                                                    & 100.0\%                                                                                & 100.0\%                                                        & 3.3\%                                                                                     \\ 
    \cline{2-8}
                                                       & OM       & 100.0\%                                                                                & 98.3\%                                                             & 18.1\%                                                                                   & 100.0\%                                                                                & 100.0\%                                                        & 22.7\%                                                                                    \\ 
    \cline{2-8}
                                                       & BPDA     & 100.0\%                                                                                & 86.9\%                                                             & 0.0\%                                                                                    & 100.0\%                                                                                & 98.2\%                                                            & 7.5\%                                                                                     \\ 
    \cline{2-8}
                                                       & SPSA     & 95.4\%                                                                                 & 89.0\%                                                             & 0.4\%                                                                                    & 90.2\%                                                                                 & 87.5\%                                                            & 9.8\%                                                                                    \\ 
    \hline
    \multirow{8}{*}{TA}                                & LLC      & 7.5\%                                                                                  & 0.6\%                                                          & 0.0\%                                                                                    & 13.2\%                                                                                 & 0.6\%                                                         & 2.1\%                                                                                     \\ 
    \cline{2-8}
                                                       & RLLC     & 2.1\%                                                                                  & 0.6\%                                                          & 0.0\%                                                                                    & 27.6\%                                                                                 & 5.8\%                                                         & 3.6\%                                                                                     \\ 
    \cline{2-8}
                                                       & ILLC     & 67.1\%                                                                                 & 19.2\%                                                         & 0.2\%                                                                                    & 100.0\%                                                                                & 99.6\%                                                        & 8.0\%                                                                                     \\ 
    \cline{2-8}
                                                       & TMIFGSM  & 84.9\%                                                                                 & 21.9\%                                                         & 1.0\%                                                                                    & 100.0\%                                                                                & 100.0\%                                                        & 27.6\%                                                                                    \\ 
    \cline{2-8}
                                                       & JSMA     & 71.0\%                                                                                 & 25.5\%                                                         & 2.5\%                                                                                    & 100.0\%                                                                                & 63.9\%                                                        & 16.1\%                                                                                    \\ 
    \cline{2-8}
                                                       & BLB      & 100.0\%                                                                                & 31.1\%                                                             & 0.7\%                                                                                    & 100.0\%                                                                                & 100.0\%                                                        & 9.2\%                                                                                     \\ 
    \cline{2-8}
                                                       & CW2      & 100.0\%                                                                                & 48.0\%                                                             & 2.8\%                                                                                    & 100.0\%                                                                                & 100.0\%                                                        & 10.4\%                                                                                    \\ 
    \cline{2-8}
                                                       & EAD      & 100.0\%                                                                                & 11.9\%                                                             & 3.0\%                                                                                    & 100.0\%                                                                                & 100.0\%                                                        & 8.7\%                                                                                     \\
    \hline
    \end{tabular}
    \caption{Attack success rate with recovered results revealed.}
    \label{tab:merged}
\end{table} 
\vspace{-15pt}
\begin{table}[H]
    \footnotesize
    \centering
    \begin{tabular}{|c|c|c|c|}
    \hline
                         & BLB  & CW2  & EAD  \\ \hline
    Original Classifier ${\mathcal C}_{\psi}$ & 0.15 & 0.17 & 0.21 \\ \hline
    With Attractor ${\mathcal M}_{(\psi, \phi)}$ & 0.41 & 0.38 & 0.45 \\ \hline
    \end{tabular}
\caption{Average $L_2$ distortion for unbounded attacks on CIFAR-10 dataset (with recovered results revealed).}
\label{tab:l2}
\end{table}
\vspace{-10pt}

\subsubsection{Unstitching Attack and Model Obfuscation}
\label{sec:obfs_result}
As our model ${\mathcal M}_{(\psi, \phi)}$ is obtained by stitching various components together, when in white-box, an adversary can study the architecture and extract the components to create custom attacks. Suppose a strong adversary is able to remove attractors completely, then the victim will be similar to an undefended model and high attack success rate can be achieved.

To prevent such attacks, obfuscation is required. We take spread spectrum watermark decoder and use method we introduced in Section~\ref{sec:obfuscation} to obfuscate the architecture of a victim model. Here we denote the obfuscated model as $\widetilde{\mathcal{M}}$. In this adaptive setting, attacker has white-box access to the victim model, the final outcome, but does not exploit analyzer's parameters.

In Table~\ref{tab:obfuscation_result}, we can see that with obfuscation, the overall attack success rate for most attacks can still be lowered significantly while having minimal impact on the overall classification accuracy. The accuracy on clean data is also not much affected by obfuscation. We have achieved $98.3\%$ and $89.1\%$ on clean samples of MNIST and CIFAR-10 respectively.

We suspect there could potentially be methods to bypass our obfuscation. However, currently we are not aware of any effective  attack that is able to unstitch this obfuscated model.
\begin{table}[H]
    \centering
    \scriptsize
    \setlength{\tabcolsep}{1.5pt}
    \begin{tabular}{|c|c|c|c|c|c|c|c|}
    \hline
                                                                      & Dataset                  & \multicolumn{3}{c|}{\textbf{MNIST}}                                                                                                                                                                                        & \multicolumn{3}{c|}{\textbf{CIFAR-10}}                                                                                                                                                                                     \\ \hline
    \begin{tabular}[c]{@{}c@{}} UA\\ /TA \end{tabular}                & Attacks                  & \begin{tabular}[c]{@{}c@{}}Initial\\Attack\\ Success\\ Rate\end{tabular} & \begin{tabular}[c]{@{}c@{}}Detection on\\Misclassified\\ Input\end{tabular} & \textbf{\begin{tabular}[c]{@{}c@{}}Overall\\Attack\\Success\\Rate\end{tabular}} & \begin{tabular}[c]{@{}c@{}}Initial\\Attack\\Success\\Rate\end{tabular} & \begin{tabular}[c]{@{}c@{}}Detection on\\ Misclassified\\ Input\end{tabular} & \textbf{\begin{tabular}[c]{@{}c@{}}Overall\\Attack\\Success\\Rate\end{tabular}} \\ \hline
    \multirow{10}{*}{UA}                                              & FGSM                     & 93.7\%                                                          & 96.9\%                                                                       & \textbf{2.9\%}                                                                    & 88.8\%                                                          & 94.6\%                                                                       & \textbf{4.8\%}                                                                     \\ \cline{2-8} 
                                                                      & RFGSM                    & 79.3\%                                                          & 96.6\%                                                                       & \textbf{2.7\%}                                                                    & 99.3\%                                                          & 95.0\%                                                                       & \textbf{5.0\%}                                                                    \\ \cline{2-8} 
                                                                      & BIM                      & 100.0\%                                                         & 75.5\%                                                                       & \textbf{24.5\%}                                                                    & 100.0\%                                                         & 98.1\%                                                                       & \textbf{1.9\%}                                                                     \\ \cline{2-8} 
                                                                      & PGD                      & 100.0\%                                                         & 75.6\%                                                                       & \textbf{24.4\%}                                                                    & 100.0\%                                                         & 97.7\%                                                                       & \textbf{2.3\%}                                                                     \\ \cline{2-8} 
                                                                      & UMIFGSM                  & 100.0\%                                                         & 79.1\%                                                                       & \textbf{20.9\%}                                                                    & 100.0\%                                                         & 97.5\%                                                                       & \textbf{2.5\%}                                                                     \\ \cline{2-8} 
                                                                      & UAP                      & 39.9\%                                                          & 99.5\%                                                                       & \textbf{0.2\%}                                                                     & 89.7\%                                                          & 99.9\%                                                                       & \textbf{0.1\%}                                                                     \\ \cline{2-8} 
                                                                      & DeepFool\footnotemark    & 100.0\%                                                         & 65.3\%                                                                       & \textbf{34.7\%}                                                                    & 91.8\%                                                          & 58.1\%                                                                       & \textbf{38.5\%}                                                                    \\ \cline{2-8} 
                                                                      & OM                       & 100.0\%                                                         & 78.2\%                                                                       & \textbf{21.8\%}                                                                    & 100.0\%                                                         & 70.1\%                                                                       & \textbf{29.9\%}                                                                    \\ \cline{2-8} 
                                                                      & BPDA                     & 100.0\%                                                         & 74.2\%                                                                       & \textbf{25.8\%}                                                                    & 100.0\%                                                         & 96.8\%                                                                       & \textbf{3.2\%}                                                                     \\ \cline{2-8} 
                                                                      & SPSA                     & 83.9\%                                                          & 99.4\%                                                                       & \textbf{0.5\%}                                                                     & 89.6\%                                                          & 100.0\%                                                                      & \textbf{0.0\%}                                                                     \\ \hline
    \multirow{8}{*}{TA}                                               & LLC                      & 11.6\%                                                          & 100.0\%                                                                      & \textbf{0.0\%}                                                                     & 10.8\%                                                          & 88.0\%                                                                       & \textbf{1.3\%}                                                                     \\ \cline{2-8} 
                                                                      & RLLC                     & 3.6\%                                                           & 100.0\%                                                                      & \textbf{0.0\%}                                                                     & 36.7\%                                                          & 71.7\%                                                                       & \textbf{10.4\%}                                                                    \\ \cline{2-8} 
                                                                      & ILLC                     & 100.0\%                                                         & 74.6\%                                                                       & \textbf{25.4\%}                                                                    & 100.0\%                                                         & 82.5\%                                                                       & \textbf{17.5\%}                                                                    \\ \cline{2-8} 
                                                                      & TMIFGSM                  & 100.0\%                                                         & 95.4\%                                                                       & \textbf{4.6\%}                                                                     & 100.0\%                                                         & 79.1\%                                                                       & \textbf{20.9\%}                                                                    \\ \cline{2-8} 
                                                                      & JSMA                     & 42.1\%                                                          & 99.0\%                                                                       & \textbf{1.0\%}                                                                     & 96.8\%                                                          & 91.5\%                                                                       & \textbf{8.2\%}                                                                     \\ \cline{2-8} 
                                                                      & BLB                      & 100.0\%                                                         & 90.3\%                                                                       & \textbf{9.7\%}                                                                     & 100.0\%                                                         & 84.3\%                                                                       & \textbf{15.7\%}                                                                    \\ \cline{2-8} 
                                                                      & CW2                      & 100.0\%                                                         & 100.0\%                                                                      & \textbf{0.0\%}                                                                     & 100.0\%                                                         & 99.0\%                                                                       & \textbf{1.0\%}                                                                     \\ \cline{2-8} 
                                                                      & EAD                      & 100.0\%                                                         & 99.9\%                                                                       & \textbf{0.1\%}                                                                     & 100.0\%                                                         & 98.3\%                                                                       & \textbf{1.7\%}                                                                     \\ \hline
    \end{tabular}
    \caption{Attack success rate on obfuscated model $\widetilde{\mathcal{M}}$.}
    \vspace{-10pt}
    \label{tab:obfuscation_result}
\end{table}
\footnotetext{Attack success rate of DeepFool can be reduced to 0\% by setting additional threshold on the difference between soft labels.}

\subsubsection{Transfer Attacks and Adversarial Training}
\label{sec:mitigate_transfer}
Previous work has shown that adversarial samples found using the substitute model could be adversarial samples to the original victim model,  that is,  adversarial samples are transferable~\cite{Papernot:2017:PBA:3052973.3053009}. 

Transfer attacks can be effective against our approach. As attackers know the details of the defense, they can bypass attractors by conducting attacks on cloned models and transfer generated adversarial samples to our model.

We consider the strongest version of transfer attacks where the substitute models have exactly same architectures as the victim models. In addition, the adversary also has huge amount of labeled data. Intuitively, such kind of attacks is hard to be defended using our approach, but we can enhance our analyzer using other defensive mechanisms. 

There are many ways to combine our method with other defenses. We give an example to combine the obfuscated model $\widetilde{\mathcal{M}}$ and an adversarially trained classifier $\mathcal{C}_{AT}$.

We place the adversarially trained classifier inside the analyzer. For a given input ${\textbf x}$, we get the prediction $\argmax \widetilde{\mathcal{M}}({\textbf x})$. We also get the prediction from the adversarially trained classifier $\argmax \mathcal{C}_{AT}({\textbf x})$. If they give different predictions, that is $\argmax \widetilde{\mathcal{M}}({\textbf x}) \neq \argmax \mathcal{C}_{AT}({\textbf x})$, we declare the input as adversarial. Otherwise, we declare the input as normal.

To see that why such combination works, note that due to the special mechanisms in adversarial training and our obfuscation method. $\widetilde{\mathcal{M}}$ and  $\mathcal{C}_{AT}$ have very different parameters. Therefore, an adversarial sample that is successful on one of them may not work on the other. Furthermore, for un-targeted attacks, even when a sample is successfully misclassified on both models, it is unlikely to be classified to the same wrong class by both models. Therefore, this combined model poses more constraints and makes searching for adversarial samples more difficult. On the other hand, as both $\widetilde{\mathcal{M}}$ and $\mathcal{C}_{AT}$ attain high accuracy on clean data, on a clean input, they are likely to give the same prediction class. Therefore, this method will not cause high false positive rate.

In this experiment, we tested the transferability of attacks on (1) victim model without any defense, (2) obfuscated model $\widetilde{\mathcal{M}}$, and (3) obfuscated model combined with an adversarially trained classifier in the analyzer $\widetilde{\mathcal{M}}_{COMB}$.

We used USPS~\cite{DBLP:journals/pami/Hull94} and CINIC-10~\cite{DBLP:journals/corr/abs-1810-03505} as the training datasets for the substitute models. They have same classes and similar distribution as MNIST and CIFAR-10 respectively.

The performance of transfer attacks are shown in Table~\ref{tab:transfer_attack}. The performance on non-transfer attacks and clean samples remains same as Section~\ref{sec:obfs_result}.
Attractors and obfuscation make the attacks slightly less transferable than on an undefended model. Placing an adversarially trained model inside the analyzer to verify the prediction greatly reduces the success rate of transfer attacks. Indeed, attractors can be used with most existing defense mechanisms to enhance the robustness.
\begin{table}[H]
    \centering
    \scriptsize
    \setlength{\tabcolsep}{2pt}
    \begin{tabular}{|c|c|c|c|c|c|c|c|}
    \hline
                                                                      & Dataset                  & \multicolumn{3}{c|}{\textbf{MNIST}}                                                                                                                                                                                                                 & \multicolumn{3}{c|}{\textbf{CIFAR-10}}                                                                                                                                                                                                              \\ \hline
    \begin{tabular}[c]{@{}c@{}}UA\\ /TA\end{tabular}                  & Attacks                  & \begin{tabular}[c]{@{}c@{}}Without\\ Defense\\ ${\mathcal C}_{\psi}$\end{tabular} & \begin{tabular}[c]{@{}c@{}}With\\ Attractor\\ +Obfus\\ $\widetilde{\mathcal{M}}$\end{tabular} & \begin{tabular}[c]{@{}c@{}}Combined\\ Model\\ $\widetilde{\mathcal{M}}_{COMB}$\end{tabular} & \begin{tabular}[c]{@{}c@{}}Without\\ Defense\\ ${\mathcal C}_{\psi}$\end{tabular} & \begin{tabular}[c]{@{}c@{}}With\\ Attractor\\ +Obfus\\ $\widetilde{\mathcal{M}}$\end{tabular} & \begin{tabular}[c]{@{}c@{}}Combined\\ Model\\ $\widetilde{\mathcal{M}}_{COMB}$\end{tabular} \\ \hline
    \multirow{10}{*}{UA}                                              & FGSM                     & 74.7\%                                                                       & 74.0\%                                                                             & 14.3\%                                                                          & 76.1\%                                                                       & 73.3\%                                                                             & 12.6\%                                                                          \\ \cline{2-8} 
                                                                      & RFGSM                    & 57.0\%                                                                       & 55.7\%                                                                             & 9.6\%                                                                           & 67.5\%                                                                       & 64.4\%                                                                             & 13.3\%                                                                          \\ \cline{2-8} 
                                                                      & BIM                      & 67.0\%                                                                       & 65.5\%                                                                             & 15.1\%                                                                          & 55.8\%                                                                       & 54.1\%                                                                             & 19.2\%                                                                          \\ \cline{2-8} 
                                                                      & PGD                      & 65.4\%                                                                       & 63.6\%                                                                             & 13.0\%                                                                          & 62.9\%                                                                       & 59.0\%                                                                             & 13.2\%                                                                          \\ \cline{2-8} 
                                                                      & UMIFGSM                  & 78.6\%                                                                       & 76.2\%                                                                             & 18.0\%                                                                          & 75.7\%                                                                       & 73.9\%                                                                             & 20.1\%                                                                          \\ \cline{2-8} 
                                                                      & UAP                      & 30.6\%                                                                       & 31.9\%                                                                             & 3.5\%                                                                           & 35.2\%                                                                       & 34.6\%                                                                             & 9.7\%                                                                           \\ \cline{2-8} 
                                                                      & DeepFool                 & 37.6\%                                                                       & 46.6\%                                                                             & 8.0\%                                                                           & 25.1\%                                                                       & 25.3\%                                                                             & 10.3\%                                                                          \\ \cline{2-8} 
                                                                      & OM                       & 40.4\%                                                                       & 49.2\%                                                                             & 7.4\%                                                                           & 26.3\%                                                                       & 26.5\%                                                                             & 10.0\%                                                                          \\ \cline{2-8} 
                                                                      & BPDA                     & 61.2\%                                                                       & 58.1\%                                                                             & 12.2\%                                                                          & 60.8\%                                                                       & 57.1\%                                                                             & 14.2\%                                                                          \\ \cline{2-8} 
                                                                      & SPSA                     & 43.7\%                                                                       & 44.5\%                                                                             & 9.1\%                                                                           & 54.6\%                                                                       & 54.0\%                                                                             & 11.4\%                                                                          \\ \hline
    \multirow{8}{*}{TA}                                               & LLC                      & 15.0\%                                                                       & 8.7\%                                                                              & 2.7\%                                                                           & 12.3\%                                                                       & 9.5\%                                                                              & 1.2\%                                                                           \\ \cline{2-8} 
                                                                      & RLLC                     & 11.1\%                                                                       & 9.2\%                                                                              & 1.4\%                                                                           & 9.1\%                                                                        & 7.2\%                                                                              & 0.8\%                                                                           \\ \cline{2-8} 
                                                                      & ILLC                     & 26.4\%                                                                       & 21.2\%                                                                             & 7.7\%                                                                           & 3.7\%                                                                        & 3.2\%                                                                              & 0.0\%                                                                           \\ \cline{2-8} 
                                                                      & TMIFGSM                  & 22.9\%                                                                       & 19.6\%                                                                             & 10.7\%                                                                          & 14.3\%                                                                       & 14.8\%                                                                             & 2.4\%                                                                           \\ \cline{2-8} 
                                                                      & JSMA                     & 8.6\%                                                                        & 8.9\%                                                                              & 1.6\%                                                                           & 4.1\%                                                                        & 3.8\%                                                                              & 1.3\%                                                                           \\ \cline{2-8} 
                                                                      & BLB                      & 16.7\%                                                                       & 15.5\%                                                                             & 9.0\%                                                                           & 3.0\%                                                                        & 3.3\%                                                                              & 1.0\%                                                                           \\ \cline{2-8} 
                                                                      & CW2                      & 4.7\%                                                                        & 6.0\%                                                                              & 1.3\%                                                                           & 3.0\%                                                                        & 3.2\%                                                                              & 0.9\%                                                                           \\ \cline{2-8} 
                                                                      & EAD                      & 0.5\%                                                                        & 0.7\%                                                                              & 0.5\%                                                                           & 2.9\%                                                                        & 2.9\%                                                                              & 0.9\%                                                                           \\ \hline
    \end{tabular}
    \caption{Transfer attack success rate on $\widetilde{\mathcal{M}}_{COMB}$.}
    \vspace{-8pt}
    \label{tab:transfer_attack}
    \end{table}

\section{Knowledgeable Attacks}
\vspace{-4pt}
While attractors are effective in confusing many known attacks, 
an important concern is whether a knowledgeable adversary, who is aware of the injection algorithm but does not know the watermarking messages,  could still bypass the attractors. In this section, we empirically evaluate two possible bypassing methods. The first method intentionally discards the soft-labels and keeps the hard-labels\footnote{Given a sample ${\textbf x}$ and a classifier ${\cal M}$,  its  hard-label is $  \argmax {\cal M}({\textbf x}) $. }, and thus does not directly utilize the soft-label gradients. The second approach smoothens the gradients so as to weaken the attractors' influences. 

\subsection{Hard-label Only Attack}
\vspace{-4pt}
Attacks which assume black-box and hard-label output are arguably less affected by attractors as they do not exploit the gradients directly. We employ Sign-OPT~\cite{DBLP:conf/iclr/ChengSCC0H20} for this evaluation. Sign-OPT is an improvement over Cheng's formulation~\cite{DBLP:conf/iclr/ChengLCZYH19} and boundary attack~\cite{DBLP:conf/iclr/BrendelRB18}. It carries out un-targeted attack  within a fixed number of queries. A large number of adversarial samples would be generated and the sample with the smallest distortion is taken as the attack's output.   By design, Sign-OPT is always successful in finding an adversarial sample. 

Although Sign-OPT  does not directly utilize the soft-label's gradient, the injected attractors could still influence  the decision boundary, which  in turn influence  local density (defined in Section~\ref{sec:choice_of_objective})  to certain extent  as demonstrated  in Figure~\ref{fig:density}. Hence, intuitively the injected attractors could provide some protections. It is interesting to determine whether the influences are sufficiently large  to  mislead the attack. 

We evaluate 1,000 test images from MNIST and CIFAR-10 respectively. Each test image is fed into Sign-OPT with  maximum number of queries set to 40,000. The found adversarial sample,  which has the smallest distortion,  is then  fed into the detection analyzer ${\mathcal F_d}$.  The analyzer's parameters are configured so that the  false positive rate (FPR)  is  $0.5\%$.  Hence, a fair comparison with other attacks  would be Table~\ref{fig:lowfpr} in Section~\ref{sec:lowfpr}.

\begin{table}[H]
    \centering
    \scriptsize
    \begin{tabular}{|c|c|c|c|} 
        \hline
        \multicolumn{2}{|c|}{Dataset}                                                                                                                                             & \textbf{MNIST}  & \textbf{CIFAR-10}   \\ 
        \hline
        \multirow{2}{*}{\begin{tabular}[c]{@{}c@{}} Average\\ Distortion \end{tabular}} & \begin{tabular}[c]{@{}c@{}}Without\\ Defense ${\mathcal C}_{\psi}$ \end{tabular}        & 1.38            & 0.102               \\ 
        \cline{2-4}
                                                                                        & \begin{tabular}[c]{@{}c@{}}With\\ Attractor ${\mathcal M}_{(\psi, \phi)}$ \end{tabular} & 2.55            & 0.0925              \\ 
        \hline
        \multicolumn{2}{|c|}{Detection Rate}                                                                                                                                      & 99.6\%          & 94.7\%              \\ 
        \hline
        \multicolumn{2}{|c|}{\textbf{Overall Attack Success Rate}}                                                                                                                         & \textbf{0.4\%}           & \textbf{5.3\%}               \\
        \hline
        \end{tabular}
    \caption{Attack success rate for Sign-OPT attack.}
    \vspace{-8pt}
    \label{tab:signopt}
\end{table}

Table~\ref{tab:signopt} shows the average distortion, detection rate and overall attack success rate.  Note that the detection analyzer is successful most of the time. In other words, the injected attractors indeed ``attract''  Sign-OPT's objective function to the attractors. 

We also modified the attack and evaluated the scenario where both detection and recovery result are exposed to the adversary. 
Recall that by design, Sign-OPT will always find an adversarial sample. Therefore, the measurement here is the average distortion. We observed that the average distortion increased from 1.38 to 8.04 for MNIST, and from 0.102 to 0.251 for CIFAR-10.

\subsection{Gradient Smoothening}
\vspace{-8pt}
In this section, we evaluate whether the attractors' influences can be effectively weakened by smoothening and  random noise. 

We design a few un-targeted attacks based on PGD (Projected Gradient Descent)~\cite{2017arXiv170606083M}. In our previous evaluation (Section~\ref{sec:eval}), PGD is one of the strongest gradient-based attacks against the proposed defense. PGD's search process randomly chooses its starting point  within a ball of interest and performs random restart, thus adds noise to its search process. Intermediate result in every iteration is clipped such that it is in the neighborhood of the original image.

Our attacks introduce additional gradient smoothening to the PGD's search process. The smoothening can be done either temporally or spatially:
\begin{itemize}[leftmargin=*]
    \item {\bf \em Momentum-based.} Instead of computing the gradient independently in each iteration, momentum (moving average of gradients) is used such that the attack follows the weighted average of gradients obtained in current and previous iterations.
    \item {\bf \em Locality-based.} Instead of using  the gradient of the given sample ${\textbf x}$ directly, a few random samples are chosen from ${\textbf x}$'s neighborhood and their average gradient is used for the search process. 
\end{itemize}
\noindent
Unlike PGD which perturbs all pixels with fixed magnitude but different signs, our attacks perturb each pixel with different magnitudes. The attack restarts and repeats for each testing sample until an adversarial sample is found. Therefore the initial attack success rate is always 100.0\%.

We evaluated these custom attacks on both un-obfuscated model ${\mathcal M}_{(\psi, \phi)}$ and obfuscated model $\widetilde{\mathcal{M}}$. We generate 1,000 adversarial samples and measure the detection rate. The false positive rate (FPR) is fixed at 0.5\%. Hence a fair comparison with previous evaluation is Table~\ref{fig:lowfpr}.

\begin{table}[H]
    \centering
    \scriptsize
    \setlength{\tabcolsep}{1.5pt}
    \begin{tabular}{|c|c|c|c|c|c|} 
        \hline
                                                                                 & Dataset                                                                                      & \multicolumn{2}{c|}{\textbf{MNIST} }                                                                                                                      & \multicolumn{2}{c|}{\textbf{CIFAR-10} }                                                                                                                     \\ 
        \hline
        Attack                                                                   & \begin{tabular}[c]{@{}c@{}} Victim\\Model \end{tabular}                                      & \begin{tabular}[c]{@{}c@{}}Detection on\\Misclassified\\ Input \end{tabular} & \textbf{\begin{tabular}[c]{@{}c@{}}Overall\\ Attack\\ Success\\ Rate \end{tabular}} & \begin{tabular}[c]{@{}c@{}}Detection on\\ Misclassified\\ Input \end{tabular} & \textbf{\begin{tabular}[c]{@{}c@{}}Overall\\ Attack\\ Success\\ Rate \end{tabular}}  \\ 
        \hline
        \multirow{2}{*}{\begin{tabular}[c]{@{}c@{}}Momentum\\Based\end{tabular}} & \begin{tabular}[c]{@{}c@{}} Un-obfuscated\\Model ${\mathcal M}_{(\psi, \phi)}$ \end{tabular} & 82.4\%                                                                       & \textbf{17.6\%}                                                            & 77.5\%                                                                        & \textbf{22.5\%}                                                             \\ 
        \cline{2-6}
                                                                                 & \begin{tabular}[c]{@{}c@{}} Obfuscated\\Model $\widetilde{\mathcal{M}}$ \end{tabular}        & 63.4\%                                                                       & \textbf{36.6\%}                                                            & 55.7\%                                                                        & \textbf{44.3\%}                                                             \\ 
        \hline
        \multirow{2}{*}{\begin{tabular}[c]{@{}c@{}}Locality\\Based\end{tabular}} & \begin{tabular}[c]{@{}c@{}}Un-obfuscated\\Model~${\mathcal M}_{(\psi, \phi)}$\end{tabular}   & 93.7\%                                                                       & \textbf{6.3\%}                                                             & 93.1\%                                                                        & \textbf{6.9\%}                                                              \\ 
        \cline{2-6}
                                                                                 & \begin{tabular}[c]{@{}c@{}}Obfuscated\\Model~$\widetilde{\mathcal{M}}$\end{tabular}          & 61.7\%                                                                       & \textbf{38.3\%}                                                            & 54.0\%                                                                        & \textbf{46.0\%}                                                             \\ 
        \hline
        \multirow{2}{*}{Combined}                                                & \begin{tabular}[c]{@{}c@{}}Un-obfuscated\\Model~${\mathcal M}_{(\psi, \phi)}$\end{tabular}   & 79.3\%                                                                       & \textbf{20.7\%}                                                            & 75.7\%                                                                        & \textbf{24.3\%}                                                             \\ 
        \cline{2-6}
                                                                                 & \begin{tabular}[c]{@{}c@{}}Obfuscated\\Model~$\widetilde{\mathcal{M}}$\end{tabular}          & 60.8\%                                                                       & \textbf{39.2\%}                                                            & 53.8\%                                                                        & \textbf{46.2\%}                                                             \\
        \hline
        \end{tabular}
    \caption{Attack success rate for custom attacks.}
    \vspace{-8pt}
    \label{tab:custom_attack}
\end{table}

In Table~\ref{tab:custom_attack}, we observe that the un-obfuscated model is still able to achieve high detection rate and significantly lower the attack success rate. The multi-scale attractors injected through varying interval sizes in QIM watermark decoder are able to misguide the adversary even when the attack attempts to target at a smoother surface. In contrast,  the obfuscated version, which implemented  spectrum decoder,  is less effective. Nonetheless, it is still able to detect more than half of the adversarial samples with low FPR. This indicates multi-scale is a desirable property.
\section{Attractors vs Trapdoors}
\label{sec:discussion}
\label{sec:relationship}
\vspace{-4pt}
Unlike an attractor, a trapdoor  $\Delta_t$ for the $t$-th class  is a global perturbation that leads almost all samples to  the  $t$-th class.  Hence, when such perturbation  $\Delta_t$ is applied to a clean sample ${\textbf x}$ that is not in  the $t$-th  class,  the perturbed sample  $({\textbf x}+\Delta_t)$ is likely to be misclassified as $t$.   

Attractors and trapdoors are related and it is possible that a model possesses properties of both. However,   there are a number of key differences and crucial implications between the two notions.
\\ \\[-7pt]
{\bf \em Implicit vs Explicit Constraints on Gradients.}\ \
The notion of trapdoor  does not explicitly impose constraints on the attack objective function. It is interesting to investigate whether the existence of a trapdoor is sufficient in misleading  the attack, so that the attack searches along the trapdoor. If this is not the case, attacks would  still be successful in the presences of trapdoors. In contrast, the notion of attractor explicitly forces the gradient to point toward the attractors. Consequently, an attack that moves along the gradient would move toward the attractors as intended.
\\ \\[-7pt]
{\bf \em Global vs Random Structure.}\ \
A trapdoor  $\Delta_t$ is ``global''  in the sense that when applied  to {\em any} sample ${\textbf x}$,  it is likely that the perturbed $({\textbf x}+\Delta_t)$ is  classified as class $t$.   On the other hand,  the notion of attractors does not dictate  that  the attractors are to be scattered  following some global properties.  This difference is analogical to the difference  of  additive watermarking vs  informed embedding watermarking~\cite{DBLP:conf/icip/MillerCB00}.  Intuitively, an additive watermark scheme adds a fixed watermark $w_0$ to any given image ${\textbf x}$, giving $({\textbf x}+w_0)$.  On the other hand,  informed embedding (e.g. QIM\cite{Chen2001})  perturbs the given image ${\textbf x}$ depending on the location  of ${\textbf x}$ in the image space,  and  thus the perturbation could be different for different images.    Such difference is crucial in watermarking since a simple averaging attack can exploit global properties of additive watermarking to derive the secret $w_0$. From this analogy, potentially there  could be averaging attacks on trapdoor that exploit its global  property.
\\ \\[-7pt]
{\bf \em Implications in Construction.}\ \
Classification model with trapdoors could be obtained through training on a mixture of the original training dataset  and  perturbed data.  More specifically, suppose  $\langle D_1, \ldots, D_n\rangle $ is the training dataset of the original victim classifier, where $D_i$ contains samples in the $i$-th class, the mixed training set is $\langle D_1 \cup \widetilde{D}_1, \ldots,  D_n \cup \widetilde{D}_n \rangle$ where $\widetilde{ D_i}$ contains samples perturbed with the trapdoor $T_i$, that is, 
$$\widetilde{D}_i = \{ \widetilde{\textbf x}\  |\  \widetilde{\textbf x}= \Delta_i +  {\textbf x},   \mbox{ where \ } {\textbf x} \in D_j   \mbox {\ for some \ } j \not= i \}$$

Shen \etal~observed that the trained model exhibits properties of the trapdoors and attains high accuracy on the original classification task. 

Here,  we argue that in an optimally trained model, the direction of the soft label's gradient might not align with the trapdoors. Hence, even if 
there are trapdoors,  we are unable to detect adversarial samples obtained from attacks. To illustrate this concern,  we give a neural network model ${\mathcal M}_{\widetilde{\theta}}$ that attains the training objective and yet cannot detect backpropagation-based adversarial attacks. 

\subsection{Construction of Trapdoor}
\label{sec:trapdoor_construct}
Our construction first obtains three neural network models ${\mathcal C}_{\pi}$, ${\mathcal D}_{\kappa}$ and ${\mathcal W}_{\omega}$ parametrized by $\pi, \kappa$ and $\omega$ respectively, and combines them to obtain ${\mathcal M}_{\widetilde{\theta}}$.  

\begin{enumerate}[leftmargin=*]
\item  ${\mathcal C}_{\pi}:{\mathcal X} \rightarrow  {\mathbb R}^n$ is the model for the original classification task. It is obtained by  training  on 
$\langle D_1, \ldots, D_n\rangle $.  We assume that the accuracy of ${\mathcal C}_{\pi}$ is high and it is difficult to further enhance its accuracy.
\item  ${\mathcal W}_{\omega}: {\mathcal X} \rightarrow  {\mathbb R}^n$ is the {\em trapdoor  decoder}, which  predicts the class of trapdoor in the input. It  can be trained on 
 $\langle \widetilde{D}_1, \ldots, \widetilde{D}_n\rangle $. While it is possible to achieve high accuracy through training,  the prediction process is 
 essentially a watermark decoding process and 
 well understood. Hence we can  analytically design an accurate  neural network classifier as the trapdoor decoder. 
 \item   ${\mathcal D_{\kappa}}:{\mathcal X} \rightarrow  {\mathbb R}$ is  the {\em trapdoor detector},  which detects whether the input contains a trapdoor.  Let us write ${\mathcal D}_{\kappa}({\textbf x})= \hat{d} $ where  $\hat{d}$ is the probability that the input contains a  trapdoor.  
 The detector can be trained on the two-class dataset $\langle D_1  \cup \ldots \cup D_n,  \    \widetilde{D}_1 \cup  \ldots \cup \widetilde{D}_n \rangle$.  Similar to  ${\mathcal W}_{\omega}$, since the detection process is well-understood, we can analytically derive an accurate  neural network classifier.
 \end{enumerate}

\begin{figure}[H]
    \vspace{-10pt}
    \begin{center}
    \includegraphics[width=0.8\linewidth]{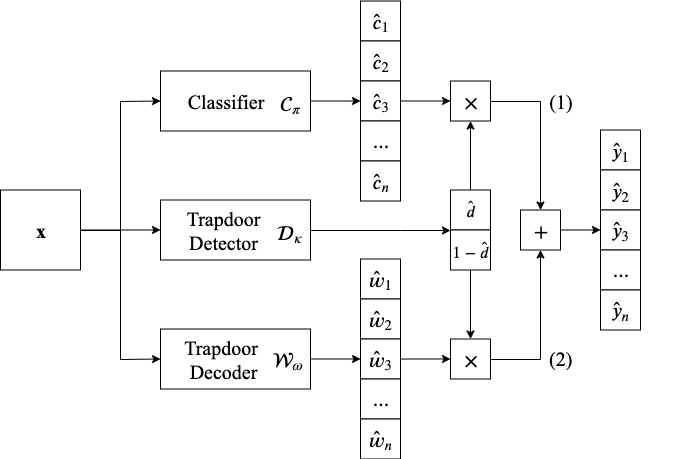}
    \end{center}
    \vspace{-10pt}
\caption{Trapdoored model meeting the training goals, but vulnerable to adversarial attacks.}
\vspace{-10pt}
\label{fig:flow_chart}
\end{figure}

Figure~\ref{fig:flow_chart} illustrates how these three models are combined.  On input ${\textbf x}$, the output of the combined model is the weighted sum:
\begin{eqnarray}
\label{eq:trapdoor}
{\mathcal M}_{\widetilde{\theta}} ({\textbf x})= {\mathcal D}_{\kappa}({\textbf x}) \cdot {\mathcal W}_{\omega}({\textbf x}) +  (1- {\mathcal D_{\kappa}}({\textbf x}))\cdot {\mathcal C}_{\pi}({\textbf x})
\end{eqnarray}
Note that the combined ${\mathcal M}_{\widetilde{\theta}}$ attains high accuracy and meets the training goals: on a clean input ${\textbf x}$ belongs to the class $t$, 
${\mathcal M}_{\widetilde{\theta}}$ behaves similar to ${\mathcal C}_{\pi}$;  on input perturbed with trapdoor,  ${\mathcal M}_{\widetilde{\theta}}$ behaves  similar to ${\mathcal W}_{\omega}$.   

Now, consider a clean input ${\textbf x}$ of class $t$, we sample in its small neighborhood and feed into  ${\mathcal D}_{\kappa}$. 
 Since ${\mathcal D}_{\kappa}$ is accurate,  the output value ${\mathcal D}_{\kappa}({\textbf x})$  would be small.  Hence, the gradients produced by
${\mathcal M}_{\widetilde{\theta}}$  would be close to the gradients by  ${\mathcal C}_{\pi}$.   Subsequently, 
a backpropagation-based attack (e.g. FGSM) on ${\mathcal M}_{\widetilde{\theta}}$  would obtain an adversarial sample ${\textbf x}'$ that is similar to the result when applied on the original victim  ${\mathcal C}_{\pi}$, and thus  the adversarial sample cannot be detected.

\subsection{Analysis of Trapdoor}
\vspace{-5pt}
We repeated the same experiment we did in Section~\ref{sec:analysis} on the trapdoored model we described in Section~\ref{sec:trapdoor_construct}. The result indeed shows that an attack conducted on the trapdoored model may not involve contribution from the trapdoors,  and thus cannot be detected by the analyzer. Details of this experiment are included in Appendix~\ref{sec:trapdoor_analysis}.
\section{Conclusion and Future Work}
\label{sec:conclusion}
\vspace{-5pt}
We have presented another angle of viewing adversarial defense mechanisms. Instead of focusing on the adversarial samples themselves, we disrupt the attacking process by injecting attractors. We propose a modular approach which provides design flexibility and explainability of the outcomes. Empirical results show that our approach can be very effective. 

We highlighted that  modular design, although provides flexibility and explainability, unfortunately also opens up another threat whereby a white-box attacker extracts out individual modules so as to aid searching of adversarial samples. This leads to a defense of obfuscating the layout and parameters. Theoretically, it could be achieved by using functional encryption or indistinguishability obfuscation. Nonetheless, such cryptographic primitives are expensive and currently there is no well-accepted candidate. Instead of using cryptographic primitives, we proposed a method that hides two components by simple summing. However, this method could not be easily extended to complex layout. Furthermore, although intuitively adversary is unable to separate the two components, thorough and rigorous formulation is required. 

In a certain way, our experiment results show that, with secure obfuscation, it is possible to attain high robustness against adversarial attacks. This leads to an interesting question on whether the fundamental difficulty in defending adversarial attack lies in obfuscation, or in some intriguing properties of the model.

\bibliographystyle{IEEEtran}
\bibliography{paper}

\newpage
\appendices
\section{Attacks Used in Experiments}
\label{sec:list_of_attacks}
In this section, we briefly describe attacks in our experiments. We denote the victim classification model as ${\mathcal M_\theta}: {\mathcal X} \rightarrow {\mathbb R}^n$ parameterized by $\theta$. Given an input ${\textbf x} \in {\mathcal X}$, ${\mathcal M_\theta}$ outputs a vector ${\hat{y}} = [\hat{y}_1\;\hat{y}_2\;\ldots\;\hat{y}_n]^T$ where each $\hat{y}_i$ represents the soft label for class $i$,  and is the probability that input ${\textbf x}$ belongs to the class $i$. The classification of ${\textbf x}$ is  the most likely  predicted class, i.e.,  $\arg\max {\mathcal M_\theta} ({\textbf x})$. We write $J(\cdot,\cdot,\cdot)$ as the classification loss function.
\\ \\[-7pt]
\noindent
{\bf \em Box-constrained L-BFGS Attack (BLB)~\cite{Szegedy2014IntriguingPO}.}\ \
Szegedy \etal~formulated generation of adversarial samples as an optimization problem. Given an input image ${\textbf x}$ and a target ${\textbf y}_{target}$, the goal is to minimize $\min \|{\textbf x}-{\textbf x}'\|_{2}$ such that ${\mathcal M_\theta}({\textbf x}) = {\textbf y}_{target}$. Szegedy \etal~transformed this goal into an easier problem: minimizing $c\cdot\|{\textbf x}-{\textbf x}'\|_{2}+J(\theta, {\textbf x}', {\textbf y}_{target})$.
\\ \\[-7pt]
\noindent
{\bf \em Fast Gradient Sign Method (FGSM)~\cite{Goodfellow2015ExplainingAH}.}\ \
FGSM moves a fixed small step $\epsilon$ in the direction that maximally changes the prediction result. The adversarial sample for an un-targeted attack is:
$$
{\textbf x}'={\textbf x}+\epsilon \cdot \operatorname{sign}\left(\nabla_{{\textbf x}} J\left(\theta, {\textbf x}, {\textbf y}_{true}\right)\right)
$$
where ${\textbf y}_{true}$ is the one-hot vector of the true label of the input ${\textbf x}$. The adversarial sample for an targeted attack is:
$$
{\textbf x}'={\textbf x}-\epsilon \cdot \operatorname{sign}\left(\nabla_{{\textbf x}} J\left(\theta, {\textbf x}, {\textbf y}_{target}\right)\right) 
$$
where ${\textbf y}_{target}$ is the one-hot vector of the target label.
\\ \\[-7pt]
\noindent
{\bf \em Basic Iterative Method (BIM)~\cite{Kurakin2017AdversarialEI}.}\ \ 
BIM is an extension of FGSM which extends the one-step attack into an iterative process. The attack chooses the starting point ${\textbf x}'_0 = {\textbf x}$ and the subsequent steps:
$$
{\textbf x}'_{n+1}=\operatorname{clip}({\textbf x}'_n+\epsilon \cdot \operatorname{sign}\left(\nabla_{{\textbf x}} J\left(\theta, {\textbf x}'_n, {\textbf y}_{true}\right)\right))
$$
\\[-7pt]
\noindent
{\bf \em Projected Gradient Descent (PGD)~\cite{2017arXiv170606083M}.}\ \
PGD is an improvement over BIM. The search process starts at a random point within the norm ball, and then follows the iterations similar to BIM.
\\ \\[-7pt]
\noindent
{\bf \em Momentum Iterative FGSM (MI-FGSM)~\cite{DBLP:journals/corr/abs-1710-06081}.}\ \ 
Dong \etal~proposed using gradients from previous iterations and applying momentum to prevent overfitting.
\\ \\[-7pt]
\noindent
{\bf \em DeepFool~\cite{MoosaviDezfooli2016DeepFoolAS}.}\ \
DeepFool finds the minimal perturbation $\epsilon$ to change the prediction result:
$$
\Delta({\textbf x} ; {\mathcal M}_{\theta}) =\min _{\epsilon}\|\epsilon\|_{2} \text { subject to } {\mathcal M}_{\theta}({\textbf x}+\epsilon) \neq {\mathcal M}_{\theta}({\textbf x})
$$
DeepFool views neural network classifiers as hyperplanes separating different classes. In a binary classifier, the minimal perturbation is the distance from ${\textbf x}_0$ to the separating hyperplane $\mathscr{M}=\left\{{\textbf x} : {\textbf w}^{T} {\textbf x}+b=0\right\}$. The minimal perturbation is the orthogonal projection of ${\textbf x}_0$ onto $\mathscr{M}$.
\\ \\[-7pt]
\noindent
{\bf \em Universal Adversarial Perturbations (UAP)~\cite{DBLP:journals/corr/Moosavi-Dezfooli16}.}\ \
Moosavi-Dezfooli \etal~proposed UAP which is a quasi-imperceptible image agnostic perturbation that can cause misclassification for most images sampled from the data distribution.
\\ \\[-7pt]
\noindent
{\bf \em OptMargin (OM)~\cite{he2018decision}.}\ \
He \etal~proposed OptMargin which generates low-distortion adversarial samples that are robust to small perturbations. This approach circumvents defenses such as transformation based defenses that sample in a small neighborhood around an input instance and get the majority prediction.
\\ \\[-7pt]
\noindent
{\bf \em Carlini and Wagner (C\&W)~\cite{towards}.}\ \
C\&W is an iterative optimization method. Its goal is to minimize the loss $\left| \epsilon\right| + c \cdot f({\textbf x}+\epsilon)$ where $f$ is an objective function such that ${\mathcal M}_{\theta}({\textbf x}+\epsilon) = {\textbf y}_{target}$ if and only if $f({\textbf x}+\epsilon) \leq 0$ and $c$ is a constant.
\\ \\[-7pt]
\noindent
{\bf \em Elastic-Net Attacks (EAD)~\cite{chen2018ead}.}\ \
EAD uses the same loss function as C\&W but combines both $L_1$ and $L_2$ penalty functions to minimize the difference between adversarial samples and original image.
\\ \\[-7pt]
\noindent
{\bf \em Least Likely Class attack (LLC)~\cite{Kurakin2017AdversarialEI}.}\ \
LLC is similar to FGSM. Instead of decreasing the score of the correct class, LLC attempts to increase the score of a least likely class. That is:
$$
{\textbf x}'={\textbf x}-\epsilon \cdot \operatorname{sign}\left(\nabla_{{\textbf x}} J\left(\theta, {\textbf x}, {\textbf y}_{LL}\right)\right)
$$
I-LLC is the iterative version of LLC.
\\ \\[-7pt]
\noindent
{\bf \em Jacobian-based Saliency Map Attack (JSMA)~\cite{Papernot2015TheLO}.}\ \
JSMA selects a few pixels in a clean sample based on the saliency map and saturates them either to the minimum or maximum value such that the new sample can be misclassified.
\\ \\[-7pt]
\noindent
{\bf \em Backward Pass Differentiable Approximation (BPDA)~\cite{athalye2018obfuscated}.}\ \
Athalye \etal~suggested that most defenses either intentionally or unintentionally break or hide the gradients as a way to prevent adversarial attack. BPDA approximates the gradient for a non-differentiable layer so that backpropagation-based attacks can be effective against such defenses.
\\ \\[-7pt]
\noindent
{\bf \em Simultaneous Perturbation Stochastic Approximation (SPSA)~\cite{pmlr-v80-uesato18a}.}\ \
SPSA uses non-gradient based optimization. By taking random small steps around the input, SPSA attempts to find the global minima.
\\ \\[-7pt]
\noindent
{\bf \em RFGSM, RLLC~\cite{tramer2017ensemble}.}\ \
Tramer \etal~proposed adding random perturbations drawn from Gaussian distribution before calculating the gradient. This targets at defenses that use gradient masking.

\section{Existing Defenses Used in Experiments}
\label{sec:list_of_defences}
In this section, we describe the three defenses we compared in our experiments.
\\ \\[-7pt]
\noindent
{\bf \em Local Intrinsic Dimensionality Based Detector (LID)~\cite{DBLP:journals/corr/abs-1801-02613}.}\ \
Intrinsic dimensionality of manifold can be seen as the minimum dimensionality required to represent a data sample on the manifold. Since data samples from a dataset can be on different manifolds, local intrinsic dimensionality (LID) is used to measure the intrinsic dimensionality of a single data sample. Ma \etal~observed adversarial perturbation can change the LID characteristics of an adversarial region. Their experiments showed adversarial samples have significantly higher estimated LID than normal samples. Based on this observation, they built a LID-based detector.
\newpage
\noindent
{\bf \em Feature Squeezing Detector (FS) ~\cite{DBLP:journals/corr/XuEQ17}.}\ \
Xu \etal~suggested the unnecessarily large feature input space often gives room for adversarial samples. They proposed feature squeezing to limit the degree of freedom for adversary. The feature squeezing methods include reduction of color depth and smoothing. The framework evaluates the prediction results of both the original input and input pre-processed by feature squeezing. The input will be identified as adversarial if the difference between any two results is larger than a certain threshold.
\\ \\[-7pt]
\noindent
{\bf \em MagNet Detector~\cite{DBLP:journals/corr/MengC17}.}\ \
Meng and Chen suggested that one of the reasons that adversarial sample can cause wrong classification is that adversarial samples are far from the normal data manifold. They built a detector that measures how different an input sample is from normal samples. Two detection mechanisms were discussed in Meng and Chen's paper: detection based on the reconstruction error of a trained autoencoder on the given dataset, detection based on probability divergence.
\section{Analysis of Trapdoor}
\label{sec:trapdoor_analysis}
\begin{figure}[H]
\centering
\begin{subfigure}{.5\linewidth}
\centering
\vspace{0pt} \includegraphics[width=0.95\linewidth]{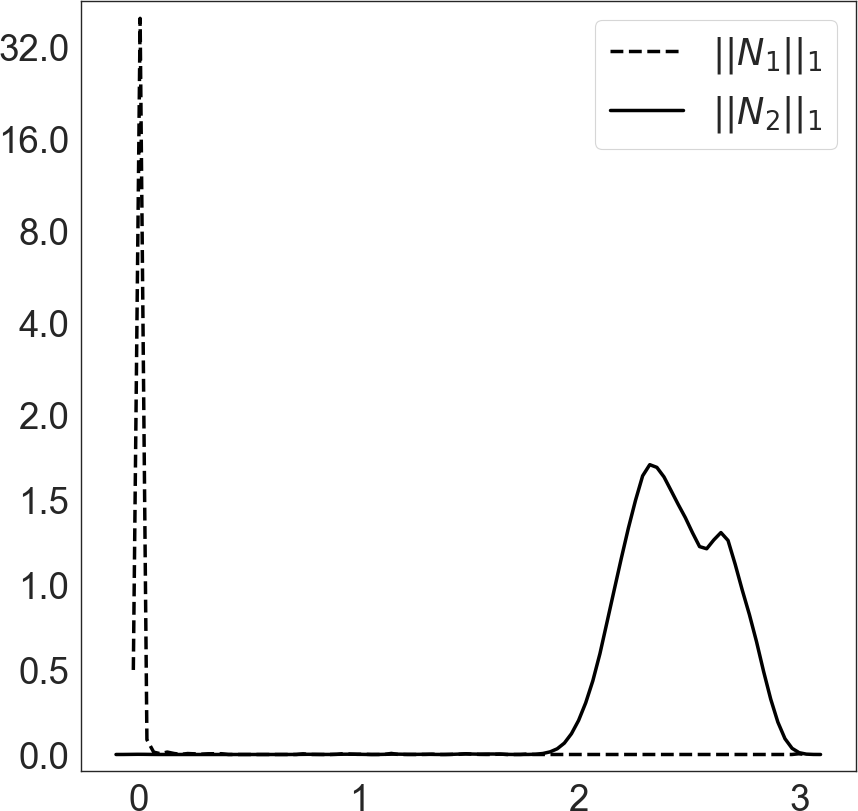}
\caption{}
\label{fig:6sub1}
\end{subfigure}%
\begin{subfigure}{.5\linewidth}
\centering
\vspace{0pt} \includegraphics[width=0.95\linewidth]{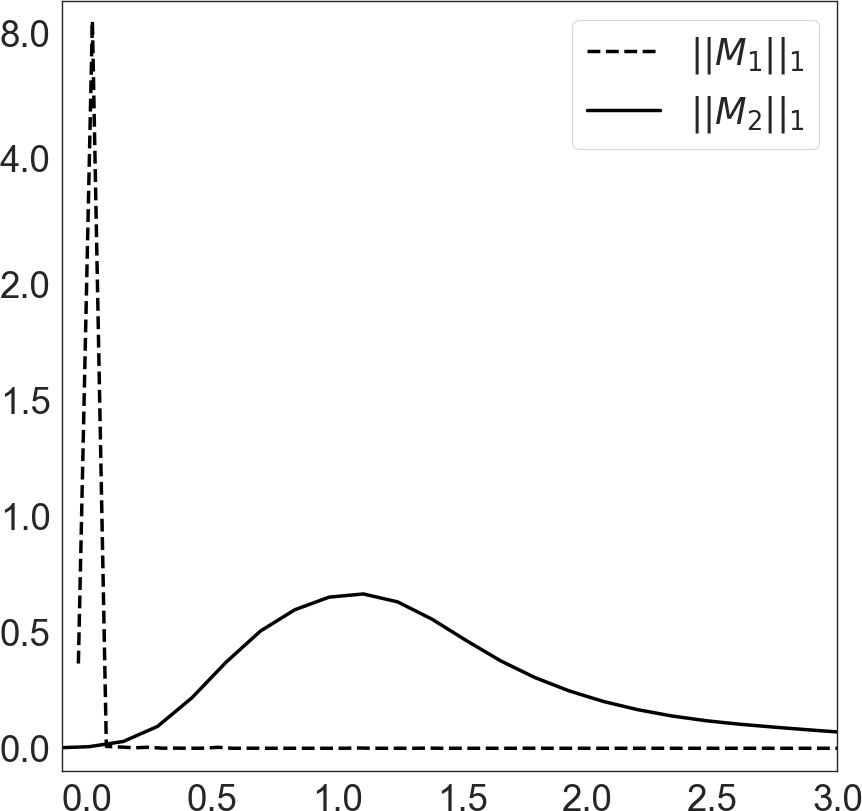}
\caption{}
\label{fig:6sub2}
\end{subfigure}
\vspace{-5pt}
\caption{Comparing  magnitude (1-norm) of outputs and gradients from ${\mathcal C}_{\pi}$  
and  $\mathcal{W}_{\omega}$. 
(a)  KDE plot of $\|N_1\|_1$ and  $\|N_2\|_1$ where $ N_1 = \mathcal{D}_{\kappa}({\textbf x}) \cdot {\mathcal W}_{\omega}({\textbf x})$ and $N_2 = (1-\mathcal{D}_{\kappa}({\textbf x})) \cdot \mathcal{C}_{\pi}({\textbf x})$. \ \  (b) KDE plot of $\|M_1\|_1$ and  $\|M_2\|_1$ where $ M_1 = \nabla_{\textbf x}{J{((\omega, \kappa), {\textbf x},{\textbf y}_{true})}}$ and $M_2 = \nabla_{\textbf x}{J{((\pi, \kappa), {\textbf x},{\textbf y}_{true})}}$.}
\label{fig:trapdoor_effect}
\end{figure}

We repeated the same experiment we did in Section~\ref{sec:analysis} on the trapdoored model we described in Section~\ref{sec:trapdoor_construct}. 
In this experiment,  the watermark decoder ${\mathcal W}_{\omega}$ is implemented based on additive watermark.
Similarly, we feed 10,000 testing images from MNIST dataset and compute the values and gradients of $N_1$ and $N_2$ for each testing image. 
Figure~\ref{fig:6sub1}  reports  the KDE derived from the measurements. 

Recall that outputs of the constructed model  is the sum of two weighted terms  given in equation (\ref{eq:trapdoor}) which corresponds to the signal at (1) and (2) in Figure~\ref{fig:flow_chart}. For convenience, let us call the two terms $N_1$ and $N_2$,  that is,   $N_1= \mathcal{D}_{\kappa}({\textbf x}) \cdot \mathcal{W}_{\omega}({\textbf x})$ and $ N_2 = (1- \mathcal{D}_{\kappa}({\textbf x}))\cdot {\mathcal C}_{\pi}({\textbf x})$ on input ${\textbf x}$.  

Figure~\ref{fig:6sub1} shows that, on clean samples,   $N_2$ dominates $N_1$, and thus the accuracy of the combined trapdoored  model would have the same accuracy as the victim model ${\mathcal C}_{\pi}$. 
 
The gradient is more complicated to determine since the term $N_1$ involves multiplication of two functions on ${\textbf x}$. 
We directly measure the two gradients using the corresponding signals (1) and (2) in Figure~\ref{fig:flow_chart}. Let us denote $(\pi, \kappa)$ the parameters of a neural network that outputs the signal (1), and $(\omega, \kappa)$ the parameters of a neural network that outputs the signal (2) in Figure~\ref{fig:flow_chart}.

Figure~\ref{fig:6sub2} shows the KDE plot of $\|\nabla_{\textbf x}{J{((\omega, \kappa), {\textbf x},{\textbf y}_{true})}\|_1}$ and $\|\nabla_{\textbf x}{J{((\pi, \kappa), {\textbf x},{\textbf y}_{true})}\|_1}$.
The gradient on clean input is  dominated by gradient of $N_2$, which is the gradient from the victim model.   Hence,  an attack conducted on the trapdoored model would  not involve contribution from the trapdoors,  and thus cannot  be detected by the analyzer. 

\section{Experiment Setup}
\label{sec:settings}
\begin{center}
    \begin{table}[H]
        \scriptsize
        \setlength{\tabcolsep}{0.0pt}
        \centering
        \begin{tabular}{|c|c|c|c|c|c|}
            \hline
             & \multicolumn{3}{c|}{\textbf{Attacks}}                    & \multicolumn{2}{c|}{\multirow{2}{*}{\textbf{Configurations}}}                              \\ \cline{2-4}
                                     & UA/TA                & Objective             & Attacks   & \multicolumn{2}{c|}{}                                                                      \\ \hline
            \multirow{16}{*}{\rotatebox[origin=c]{90}{\textbf{MNIST}}}  & \multirow{8}{*}{UAs} & \multirow{6}{*}{$L_{\infty}, \epsilon=0.3$} & FGSM      & \multicolumn{2}{c|}{$\epsilon=0.3$}                                                                   \\ \cline{4-6} 
                                     &                      &                       & RFGSM    & $\epsilon=0.15$           & $\alpha=0.15$                                                                     \\ \cline{4-6} 
                                     &                      &                       & BIM       & $\epsilon=0.3$               & \begin{tabular}[c]{@{}c@{}}iteration=15;\\ ieps\_iter=0.05\end{tabular}                                                             \\ \cline{4-6} 
                                     &                      &                       & PGD       & $\epsilon=0.3$               & \begin{tabular}[c]{@{}c@{}}iteration=15;\\ ieps\_iter=0.05\end{tabular}                                                              \\ \cline{4-6} 
                                     &                      &                       & UMIFGSM & $\epsilon=0.3$               & \begin{tabular}[c]{@{}c@{}}iteration=15;\\ ieps\_iter=0.05\end{tabular}                                                             \\\cline{4-6} 
                                     &                      &                       & UAP & $\epsilon=0.3$               & fool rate=30\%                                                            \\ \cline{3-6} 

                                     &                      & \multirow{2}{*}{$L_2$}                    & DF        & overshoot=0.02 & max\_iter=50                                                               \\ \cline{4-6}
                                    &                      &                       & OM & {\begin{tabular}[c]{@{}l@{}}batch size=1000;\\ initial const=0.02;\\ bin search steps=4\end{tabular}} & {\begin{tabular}[c]{@{}l@{}}learning rate=0.2;\\ noise count=20;\\ noise mag=0.3\end{tabular}} \\ \cline{2-6} 
                                     & \multirow{8}{*}{TAs} & \multirow{4}{*}{$L_{\infty}, \epsilon=0.3$} & LLC       & \multicolumn{2}{c|}{$\epsilon=0.3$}                                                                      \\ \cline{4-6} 
                                     &                      &                       & RLLC     & $\epsilon=0.15$           & $\alpha=0.15$                                                                           \\ \cline{4-6} 
                                     &                      &                       & ILLC      & $\epsilon=0.3$               & eps\_iter=0.05                                                             \\ \cline{4-6} 
                                     &                      &                       & TMIFGSM & $\epsilon=0.3$               & eps\_iter=0.05                                                             \\ \cline{3-6} 
                                     &                      & $L_0$                    & JSMA      & $\theta=1$               & $\gamma=0.1 $                                                                         \\ \cline{3-6} 
                                     &                      & \multirow{3}{*}{$L_2$}   & BLB       &binary\_step=5 & \begin{tabular}[c]{@{}c@{}}maximum\_iteration=1000;\\ init const=0.01\end{tabular}                                                                   \\ \cline{4-6} 
                                     &                      &                       & CW2       & \begin{tabular}[c]{@{}c@{}}batch\_size=10;\\ learning\_rate=0.02\end{tabular}               & \begin{tabular}[c]{@{}c@{}}maximum\_iteration=10000;\\ init const=0.001 ; \\ box=-0.5, 0.5\end{tabular}  \\ \cline{4-6} 
                                     &                      &                       & EAD       & \begin{tabular}[c]{@{}c@{}}binary\_step=10;\\ learning\_rate=0.02\end{tabular}               & \begin{tabular}[c]{@{}c@{}}maximum\_iteration=10000;\\ $\beta=1e-3$\end{tabular}                                                                             \\ \hline

            \multirow{16}{*}{\rotatebox[origin=c]{90}{\textbf{CIFAR-10}}}  & \multirow{8}{*}{UAs} & \multirow{6}{*}{$L_{\infty}, \epsilon=0.1$} & FGSM      & \multicolumn{2}{c|}{$\epsilon=0.1$}                                                                   \\ \cline{4-6} 
                                     &                      &                       & RFGSM    & $\epsilon=0.05$           & $\alpha=0.05$                                                                     \\ \cline{4-6} 
                                     &                      &                       & BIM       & $\epsilon=0.1$               & \begin{tabular}[c]{@{}c@{}}iteration=15;\\ ieps\_iter=0.01\end{tabular}                                                             \\ \cline{4-6} 
                                     &                      &                       & PGD       & $\epsilon=0.1$               & \begin{tabular}[c]{@{}c@{}}iteration=15;\\ ieps\_iter=0.01\end{tabular}                                                            \\ \cline{4-6} 
                                     &                      &                       & UMIFGSM & $\epsilon=0.1$               & \begin{tabular}[c]{@{}c@{}}iteration=15;\\ ieps\_iter=0.01\end{tabular}                                                             \\ \cline{4-6}  
                                     &                      &                       & UAP & $\epsilon=0.1$               & fool rate=30\%                                                             \\ \cline{3-6} 
                                     &                      & \multirow{2}{*}{$L_2$}                    & DF        & overshoot=0.02 & max\_iter=50                                                               \\ \cline{4-6}
                                    &                      &                       & OM & {\begin{tabular}[c]{@{}l@{}}batch size=1;\\ initial const=1;\\ bin search steps=4\end{tabular}} & {\begin{tabular}[c]{@{}l@{}}learning rate=0.02;\\ noise count=20;\\ noise mag=0.3\end{tabular}} \\ \cline{2-6}
                                     & \multirow{8}{*}{TAs} & \multirow{4}{*}{$L_{\infty}, \epsilon=0.1$} & LLC       & \multicolumn{2}{c|}{$\epsilon=0.1$}                                                                      \\ \cline{4-6} 
                                     &                      &                       & RLLC     & $\epsilon=0.05$           & $\alpha=0.05$                                                                           \\ \cline{4-6} 
                                     &                      &                       & ILLC      & $\epsilon=0.1$               & eps\_iter=0.01                                                             \\ \cline{4-6} 
                                     &                      &                       & TMIFGSM & $\epsilon=0.1$               & eps\_iter=0.01                                                             \\ \cline{3-6} 
                                     &                      & $L_0$                    & JSMA      & $\theta=1$               & $\gamma=0.1 $                                                                         \\ \cline{3-6} 
                                     &                      & \multirow{3}{*}{$L_2$}   & BLB       &binary\_step=5 & \begin{tabular}[c]{@{}c@{}}maximum\_iteration=1000;\\ init const=0.01\end{tabular}                                                                 \\ \cline{4-6} 
                                     &                      &                       & CW2       & \begin{tabular}[c]{@{}c@{}}binary\_step=10;\\ learning\_rate=0.02\end{tabular}               & \begin{tabular}[c]{@{}c@{}}maximum\_iteration=10000;\\ init const=0.001; \\ box=-0.5, 0.5\end{tabular} \\ \cline{4-6} 
                                     &                      &                       & EAD       & \begin{tabular}[c]{@{}c@{}}binary\_step=10;\\ learning\_rate=0.02\end{tabular}                & \begin{tabular}[c]{@{}c@{}}maximum\_iteration=10000;\\ $\beta=1e-3$ \end{tabular}                                                                            \\ \hline
            \end{tabular}
            \caption{Settings of adversarial attacks used in comparison with LID, FS and MagNet. Table is from DEEPSEC~\cite{Ling2019DEEPSECAU}.}
            \label{table:setup}
    \end{table}
\end{center}

\begin{table}[H]
    \scriptsize
    \setlength{\tabcolsep}{2.5pt}
    \centering
    \begin{tabular}{|c|c|c|c|}
    \hline
    UA/TA                    & Attacks & \multicolumn{2}{c|}{Attack Configuration} \\ \hline
    \multirow{4}{*}{UAs} & FGSM          &  \multicolumn{2}{c|}{$\epsilon=0.3$}                    \\ \cline{2-4} 
                        & PGD           & $\epsilon=0.3$                    & \begin{tabular}[c]{@{}c@{}}iteration=100;\\ eps\_iter=0.04\end{tabular}                    \\ \cline{2-4} 
                        & BPDA          & $\epsilon=0.3$                    & \begin{tabular}[c]{@{}c@{}}iteration=100;\\ eps\_iter=0.04\end{tabular}                     \\ \cline{2-4} 
                        & SPSA         &\begin{tabular}[c]{@{}c@{}}$\epsilon=0.3$;\\ learning\_rate=0.1\end{tabular}                    & iteration=500                     \\ \hline
    \multirow{2}{*}{TAs} & CW2            & \begin{tabular}[c]{@{}c@{}}binary\_step=20;\\ learning\_rate=0.1\end{tabular}                    & \begin{tabular}[c]{@{}c@{}}maximum\_iteration=500;\\ init const=0.001\end{tabular}                    \\ \cline{2-4} 
                        & EAD           &\begin{tabular}[c]{@{}c@{}}binary\_step=20;\\ learning\_rate=0.5\end{tabular}                     & \begin{tabular}[c]{@{}c@{}}maximum\_iteration=500;\\ init const=0.001\end{tabular}                    \\ \hline
    \end{tabular}
    \caption{Settings of adversarial attacks used in comparison with Trapdoor. Table is from Trapdoor paper~\cite{DBLP:journals/corr/abs-1904-08554}.}
    \label{tab:setting_trapdoor}
\end{table}

\end{document}